\documentclass[11pt]{article}
\usepackage{epsfig,amsfonts}
\usepackage[fleqn]{amsmath}
\usepackage{amsthm,amssymb}
\usepackage{graphicx}
\usepackage{hhline}
\usepackage{cite}
\usepackage{wasysym}
\usepackage{mathrsfs}
\def\be{\begin{equation}}
\def\ee{\end{equation}}
\def\bdm{\begin{displaymath}}
\def\edm{\end{displaymath}}
\def\bea{\begin{eqnarray}}
\def\eea{\end{eqnarray}}

\newcommand{\rd}{\mbox{d}}
\newcommand{\ri}{\mbox{i}}
\newcommand{\re}{\mbox{e}}

\begin{document}
\begin{titlepage}
\begin{flushright}
LPTA-05-58\\
\end{flushright}

\vspace{0.3cm}

\begin{center}
\begin{LARGE}
{\bf  Boundary  RG Flow Associated}

\vspace{0.3cm}
{\bf with  the AKNS Soliton Hierarchy}

\end{LARGE}

\vspace{1.3cm}

\begin{large}

{\bf Vladimir A. Fateev}$^{1,3}$ {\bf and Sergei  L. Lukyanov}$^{2,3}$
\end{large}

\vspace{1.cm}

{${}^{1}$ Laboratoire de Physique Th${\acute {\rm e}}$orique et Astroparticules\\
Universit${\acute {\rm e}}$ Montpellier II\\
Pl. E. Bataillon, 34095 Montpellier,
France\\

\vspace{.2cm}

${}^{2}$NHETC, Department of Physics and Astronomy\\
     Rutgers University\\
     Piscataway, NJ 08855-0849, USA\\

\vspace{.2cm}

and\\

\vspace{.2cm}
${}^{3}$L.D. Landau Institute for Theoretical Physics\\
  Chernogolovka, 142432, Russia\\
}
\vspace{1.5cm}

\end{center}

\begin{center}
\centerline{\bf Abstract} \vspace{.8cm}
\parbox{11cm}{
We introduce and study  an integrable boundary flow possessing an
infinite number of conserving charges  which  can be thought of as
quantum counterparts of  the  Ablowitz, Kaup, Newell and Segur
Hamiltonians.
We propose an exact expression for overlap
amplitudes of the boundary state with all primary states in terms
of solutions of certain ordinary linear differential equation.
The boundary flow is terminated at a nontrivial infrared fixed point.
We  identify a form of whole boundary state corresponding
to this fixed point.
 }
\end{center}

\begin{flushleft}
\rule{4.1 in}{.007 in}\\
{October 2005}
\end{flushleft}
\vfill
\end{titlepage}
\newpage

\tableofcontents

\section{ Introduction} \label{secintro}

The so called hairpin  model of boundary interaction was introduced in
\cite{LVZ}. This  two-dimensional model of Quantum Field Theory (QFT)
involves a two-component Bose field
${\bf X}(\sigma,\tau)= \big(X(\sigma,\tau)
, Y(\sigma,\tau)\big)$ which lives on the semi-infinite
cylinder $\tau\equiv\tau+2\pi R,\ \sigma\geq 0$.
In the bulk, $\sigma>0$, the field ${\bf X}$ is a free
massless field, as described by the bulk action
\bea\label{baction}
{\mathscr   A}_{\rm bulk}=
{1\over \pi}\ \int_0^{2\pi R}\rd\tau\int_{0}^{\infty}\rd\sigma\
\partial{\bf  X}\cdot{\bar \partial}{\bf  X}\ ,
\eea
with
$\partial={1\over 2}\ (\partial_{\sigma}-\ri\, \partial_{\tau})$
and $
{\bar \partial}={1\over 2}\ (\partial_{\sigma}+\ri\, \partial_{\tau})$.
The boundary values  of this field, $ {\bf
  X}|_{\sigma=0}=(X_B,\, Y_B)$, are subjected to a nonlinear constraint
\bea\label{bconstaint} \exp\big(
{\textstyle{X_B\over\sqrt{n}}}\big) - \cos\big(
{\textstyle{Y_B\over\sqrt{n+2}}}\big)= 0 \ , \eea where   $n$ is a
positive parameters. A remarkable feature of the model is that it
possesses an extended conformal symmetry with respect to certain
$W$-algebra. The  generating  holomorphic, $W_{s}=
W_s(\sigma+\ri\tau)$, and   antiholomorphic, ${\bar W}_s={\bar
W}_s(\sigma-\ri\tau)$, currents of this algebra have spins $s=2,\,
3,\, 4\ldots$ and  $s=-2,\, -3,\, -4\ldots$ respectively. Among
them there are  spin-($\pm 2)$ currents which are components of
the stress-energy tensor: \bea\label{energymom} W_2 &=& -
\partial  X \partial X - \partial Y \partial Y +
\textstyle{1\over\sqrt{n}}\,\partial^2 X\,,\\ \nonumber {\bar W}_2
&=& - {\bar \partial} X
 {\bar \partial} X - {\bar \partial} Y {\bar \partial} Y +
\textstyle{{1\over\sqrt{n}}}\,{\bar \partial}^2 X\,  .
\eea
The first nontrivial holomorphic current
reads explicitly as follows
\bea\label{wwwcurrent}
\nonumber
W_3 &=& {\textstyle {6n+4\over 3}}\ \big(\partial Y\big)^3 +
2n\ \big(\partial X\big)^2 \partial Y +\\
&& n\sqrt{n}\ \partial^2
X\partial Y - (n+2)\sqrt{n}
\ \partial X \partial^2 Y +
{\textstyle {{n+2}\over 6}}\ \partial^3 Y\ ,
\eea
while the higher currents $W_4,\, W_5\ldots$ can be generated recursively
from the singular parts of
operator product expansions of the lower currents.
The   antiholomorphic currents ${\bar W}_s$ can be
obtained from the corresponding
holomorphic one  by means of the
formal substitution $\partial\to{\bar\partial}$.

As usual in QFT, an effect of the boundary can be described in
terms of the boundary state which incorporates all information
about  boundary conditions\ \cite{nappi,Ishibashi,Cardy,gz}. In
our case the boundary state $|\, B\, \rangle_{\rm hair}$ is a
special vector in the space of states ${\cal H}$ of the
two-component uncompactified scalar associated with the ``equal
time section'' $\sigma=const$: \bea\label{ldkd} |\, B\,
\rangle_{\rm hair}\in {\cal H}=\int_{{\bf P}}{\cal F}_{\bf
P}\otimes {\bar {\cal F}}_{ {\bf P}}\ , \eea where ${\cal F}_{\bf
P}\ ({\bar {\cal F}}_{\bf  P})$ is the Fock space of two-component
right-moving (left-moving) boson with the zero-mode momentum ${\bf
P}=(P,\,Q)$. The above mentioned $W$-invariance of the  boundary
condition\ \eqref{bconstaint}\ implies that the corresponding
boundary state obeys an infinite set of equations\
\cite{Ishibashi,Cardy}: \bea\label{constrW} \big[\, W_{s+1}(\tau)-
{\bar W}_{s+1}(\tau)\, \big]_{\sigma=0}\, |\, B\, \rangle_{\rm
hair}=0\ . \eea

Once preserves the conformal symmetry, the hairpin boundary
condition can be treated as a Renormalization Group (RG) fixed
point in the space of  boundary interactions of two-component free
Bose field. Broadly speaking any  relevant boundary perturbation
breaks down the scale invariance of the original  model and
introduces some RG invariant ``physical scale'' $E_*$ in the
theory. Unfortunately there is no systematic machinery for study
an arbitrary    perturbation. Therefore it makes sense   to
consider a particular class of   perturbations preserving  some
amount of the original $W$-symmetry. The boundary state for
such  models satisfies the conditions \bea\label{alksla} (\,
{\mathbb I}_s- {\bar {\mathbb I}}_s\,)\ |\, B\, \rangle_{\rm
pert}=0 \eea for some operator-valued functionals ${\mathbb I}_s$
$({\bar {\mathbb I}_s}$)  of the original holomorphic
(antiholomorphic)  $W$-currents of the form \bea\label{aks}
{\mathbb I}_s=\int_0^{2\pi R}{\rd\tau\over 2\pi}\ P_{s+1}\ , \ \ \
\ \ \ \ {\bar {\mathbb I}}_s= \int_0^{2\pi R}{\rd\tau\over 2\pi}\
{\bar P}_{s+1}\ , \eea where the densities
$P_{s+1}=P_{s+1}[W_2,W_3\ldots] , \, {\bar P}_{s+1}=P_{s+1}[{\bar
W}_2,{\bar W}_3\ldots]$ are appropriately regularized polynomials
in $W$-currents and their derivatives. Here the subscript $s+1$
labels the spin of the local field $P_{s+1}$. Roughly speaking,
the meaning of \eqref{alksla} is that the boundary neither emits
nor absorbs any amount of the combined charges ${\bar {\mathbb
I}}_s- {\mathbb I}_s$. For this reason we shall call ${\mathbb
I}_s$ as a local Integrals of Motion (IM) of spin $s$. Notice that
in the case
 \bea\label{alsjs} P_2= W_2\, , \eea
 where $W_2$ is the holomorphic component of stress-energy
 tensor, Eq.\eqref{alksla}
manifests the invariance with respect to translations along the
$\tau$-direction.

Let us  assume that an infinite sequence of polynomials\ $\{\,
P_{s+1}\, \}$, such that the associated IM are mutually
commutative, \bea\label{lksja} [\, {\mathbb I}_s\, ,\, {\mathbb
I}_{s'}]=0\ , \eea is given. It is natural to  expect that  a
theory possessing  such an infinite commuting  set is integrable
\cite{gz}.

At the best of our knowledge the
complete algebraic classification
of  infinite commuting  sets
of local IM for  the hairpin $W$-algebra
has not been obtained yet.
Nevertheless at least
three nontrivial examples  are known\ \cite{FATT,VFat,VVFat}.
In Ref.\cite{LVZ}\ it was studied  the RG boundary flow
associated with the
so called ``paperclip series'' of local IM. The series
contains
local IM with the odd spins $s=1,\, 3,\, 5\ldots$\ .
The  second known Abelian  subalgebra\ \cite{VFat}
is deeply  related to the  Ablowitz, Kaup, Newell and Segur
(AKNS) soliton hierarchy\ \cite{Zahar,AKNSH}.
In fact the corresponding  ${\mathbb I}_s$
are
the quantum counterparts  of
the AKNS Hamiltonians.
For this reason we shall refer
to this infinite sequence of commuting integrals
as AKNS series.
Among  characteristic properties
of this series is that it contains
the local IM
with
$s=1,\, 2,\, 3\ldots$ and
the first two local densities
are  given by Eq.\eqref{alsjs}\ and\
\bea\label{lkdpao}
P_3=\ri\ W_3\ .
\eea
One more  series of
the local IM containing
the odd spins only is known. Despite
admissible spins of ${\mathbb I}_s$ for this series
and for the paperclip series are the same,
they are not equivalent.
An explicit form of  ${\mathbb I}_3$
from  the third
series can be found in \cite{FATT}.

In this article we  study an integrable model associated with the
AKNS series  of local IM. The  theory  can be defined by adding
some special boundary term  of the form \bea\label{perturb}
{\mathscr   A}_{\rm pert}= \int_0^{2\pi R}{\rd\tau\over 2\pi}\
U({\bf X}_B) \eea to the bulk action \eqref{baction}. The
potential $U({\bf X}_B)$   turns out  to be unbounded and pure
imaginary. In spite of these  somewhat pathological properties, the
corresponding  QFT appears to be well defined and possesses many
remarkable features. We shall refer to this theory as the
Integrable Perturbed Hairpin (IPH)  model\footnote{Because of the
unbound property of $U({\bf X}_B)$ the term ``perturbed'', here
and bellow, is used in   a    loose sense.}.

As was pointed out  in Ref.\cite{blz} a  boundary state associated
with integrable boundary flows with   conformal bulk can be
studied in a framework of Quantum Inverse Scattering method\
\cite{SKLYN,FST,Korep}. To   recall  the basic idea by the example
of IPH model let us assume
that the corresponding local IM are hermitian operators and the
set   $\{\, {\mathbb I}_s\, \}_{s=1}^{\infty}$
is  ``resolving'',
i.e., that all
eigenspaces of the local IM are one-dimensional
and mutually orthogonal (which seems to be
the case, see Section\, \ref{secfour}).
Then, as it follows from the structure of space ${\cal H}$\ \eqref{ldkd}
and  the condition
\eqref{alksla},\ the IPH boundary state
can be written as
\bea\label{ssshsaksa}
|\, B\, \rangle_{\rm {iph}}=\int_{\bf P}\rd^2{\bf P}\, \sum_{\alpha}
B_{\alpha}({\bf P})\ |\, \alpha, {\bf P}\, \rangle\otimes
{\overline {|\, \alpha, {\bf P}\, \rangle}}\ ,
\eea
where  $\{|\, \alpha,\, {\bf P}\, \rangle\}$
is the orthonormalized basis of
eigenvectors   in each Fock space ${\cal F}_{\bf P}$ labeled by some  index
$\alpha$.
In a  view of  Eq.\eqref{ssshsaksa},
it is convenient to think of the
boundary state in terms of the associated {\it boundary operator}.
The natural isomorphism  between ${\cal F}_{{\bf P}}$ and ${\bar
{\cal F}}_{{\bf P}}$ (the right movers are replaced by the left
movers) makes it possible to establish one to one correspondence
between states in ${\cal F}_{{\bf P}}\otimes{\bar {\cal F}}_{{\bf
P}}$ and operators in ${\cal F}_{{\bf P}}$. Thus the boundary
state $|\, B\, \rangle_{\rm iph}$ can be reinterpreted as an operator
\bea\label{ytkajsh}{\mathbb B}=
\int_{\bf P}\rd^2{\bf P}\,\sum_{\alpha}
B_{\alpha}({\bf P})\ |\, \alpha, {\bf P}\, \rangle\langle\, \alpha, {\bf P}\,|\ ,
\eea
which  commutes with the all local IM:
\bea\label{assjs} [\, {\mathbb B}\, ,\, {\mathbb I}_s\, ]=0\ .
\eea

The structure \eqref{ssshsaksa}\ emphasizes an importance of the
problem of simultaneous diagonalization of local IM ${\mathbb
I}_s$ as operators acting in  ${\cal F}_{\bf P}$. In this
connection, it is pertinent to remind that physical quantities
like the boundary state (operator) essentially  depend on some RG
invariant scale $E_*$. For the IPH model  it is convenient to
choose this dependence in the form ${\mathbb B}={\mathbb
B}(\lambda)$, where the dimensionless parameter $\lambda$ is
related with this scale by
\bea\label{kajshk} \lambda= \Big({ E_*
R\over n}\Big)^{n+2\over n}\ . \eea At the same time the
associated local IM do not involve any particular energy scale and
their eigenvectors, $\{|\, \alpha,\, {\bf P}\, \rangle\}$,
do  not depend on  $\lambda$. Hence
the  operators ${\mathbb B}$\ \eqref{ytkajsh}
corresponding to different values of the ``spectral'' parameter
$\lambda$ commute between themselves: \bea\label{kjhaksj} [\,
{\mathbb B}(\lambda)\, ,\, {\mathbb B}(\lambda')\, ]=0\ . \eea

The problem of simultaneous diagonalization of commuting operator
families is a typical problem for  the Quantum Inverse Scattering
method\ \cite{SKLYN,FST,Korep}. In this approach  commuting families,
like $ {\mathbb B}(\lambda)$\ \eqref{kjhaksj}, are defined in
terms of certain monodromy matrices associated with an  auxiliary
linear problem  where $\lambda$ plays a  role of  spectral
parameter. In Section\,\ref{secone}, we briefly discuss   the  IPH
model in the classical limit to clarify its relation to the AKNS
soliton hierarchy. Later, in Section\ \ref{seceight2}, we shall
present  arguments that the  boundary operator $ {\mathbb
B}(\lambda)$ in the quantum theory can be treated as a version of
Baxter's $Q$-operator\ \cite{Baxter}. In particular, it satisfies
the famous Baxter $T-Q$ equation: \bea\label{aksks} {\mathbb
B}(\lambda)\, {\mathbb T}(\lambda)= {\mathbb B}\big(\lambda {\bf
q}\big) + {\mathbb B}\big(\lambda {\bf q}^{-1}\big)\ \ \ \ \
\big(\, {\bf q}=\re^{-{2\pi{\rm i}\over n}}\, \big)\ , \eea
where
the transfer-matrix ${\mathbb T}(\lambda)$ is a trace of
quantum $2\times 2$ monodromy matrix for the    auxiliary AKNS
linear problem. The operator ${\mathbb T}(\lambda)$ can be thought
of as a  generating function for the AKNS series of local IM.

It should be emphasized that the paper does not contain a rigorous
quantization procedure of the AKNS  hierarchy. It  is devoted to
study of the simplest  boundary amplitude \bea\label{lksjsa}
Z={}_{\rm{iph}}\langle\, B\, |\, {\bf P}\,  \rangle\ , \eea where
$|\, {\bf P}\, \rangle\in {\cal H}$ is the highest vector in the
Fock modulus ${\cal F}_{\bf P}\otimes {\bar {\cal F}}_{\bf  P}$
corresponding to an arbitrary ${\bf P}$. A wealth of data about
$Z$ can be obtained through perturbative analysis in the weak
coupling domain, and by looking into various limiting cases of the
model; Sections\ \ref{sectwo}-\ref{secsix} are devoted to these
tasks. Using these data we propose in Section\ \ref{secseven} an
exact expression for the vacuum overlap \eqref{lksjsa} in terms of
solutions of  certain ordinary differential equation. Only in
Section\ \ref{seceight}, examining  properties of the vacuum
amplitude $Z$, we reveal  general integrable structures, like
Baxter $T-Q$ equation\ \eqref{aksks}, inherent in the theory. We
conclude the paper with  a discussion of   an  infrared fixed
point of the  boundary flow.

\section{The classical  IPH model}
 \label{secone}

Before going over to QFT, we will
explore the classical limit of the model.
For this purpose it is convenient to use the field
\bea\label{alsjlj}
{\bf x}=(x,y)={{\bf X}\over \sqrt{n}}\ .
\eea
Indeed, if one rewrites\ \eqref{baction}\ in terms of ${\bf x}$
the parameter $n$  appears  in-front of the Gaussian action.
This allows one
to interpret ${n\over 2\pi}$ as the inverse Plank constant.

\subsection{Integrable perturbation of the classical
hairpin model}

As $n\to \infty$,
$x$ and $y$ become classical  fields subjected by
the boundary condition
\bea\label{alsjal}
\big[\,  \re^{x}-\cos(y)\, \big]_{\sigma=0}=0\ .
\eea
The classical equations
of motion in  the unperturbed hairpin model
include the bulk equations, $\triangle{ x}=\triangle{ y}=0$,
as well as  the  boundary equation
\bea\label{kaash}
\partial_{\sigma}{\bf x}\cdot {\bf t}\, |_{\sigma=0}=0\ ,
\eea where ${\bf t}=( -\tan(y), 1)$ is a tangent vector to the  curve\
\eqref{alsjal}. To take into account  the zero mode, we shall
consider the classical solutions ${\bf x}(\sigma,\tau)$ such that
\bea\label{saklskh} {\bf x}(\sigma,\tau)\to {2\ri
{\boldsymbol\xi}\over R}\ \ \sigma\ \ \ \ \ \ {\rm as}\ \ \ \
\sigma\to+\infty\ , \eea where ${\boldsymbol\xi}=(\xi_x,\, \xi_y)$
is  some constant vector.

In the large $n$
limit the quantum  field $W_3$\ \eqref{wwwcurrent}\ produces
a classical holomorphic current,
$W_3\to n^{5\over 2}\  w_3$,
with
\bea\label{amsbj}
w_3=2\, (\partial y)^3+2\, (\partial x)^2\, \partial y+
\partial^2 x\,\partial y-\partial^2 y\,\partial x
\  ,
\eea
and Eq.\eqref{constrW} for $s=3$ implies
that the difference $w_3-{\bar w}_3$ vanishes
at the boundary. One can indeed check
that the last condition  holds in virtue of
the classical equations of
motion
and the boundary constraint\ \eqref{alsjal}.

Now let us analyze   an effect of the
boundary potential\ \eqref{perturb} in the classical theory.
We
still assume the boundary constraint\ \eqref{alsjal}, so the perturbation
modifies the classical
boundary equation of motion\ \eqref{kaash}\ only:
\bea\label{lksj}
 \partial_{\sigma}{\bf x}\cdot {\bf t}\, |_{\sigma=0}=f\ ,
\eea where $f={1\over \sqrt{n}} \ \nabla_{{\bf X}_B} U\cdot{\bf
t}$. For an arbitrary function  $U({\bf X}_B)$ the holomorphic
current $w_3$ does not generate a conserving charge because
\bea\label{askas} \big[\, w_3-{\bar w}_3\, \big]_{\sigma=0}&=&
{\textstyle {\ri \over 2}}\ \Big[ -{\textstyle {{\rm d}\over {\rm
d}\tau}} \big(f(x)
\partial_{\sigma} x\big)+3f^2 \cot(y)\,
\partial_{\tau}x+\nonumber\\ &&2\, \big(\, {\textstyle{{\rm d}
f\over {\rm d} x}} + 2f\, \big)\,
\partial_{\tau}x\partial_{\sigma}x \,
 \Big]_{\sigma=0}\ .
\eea However, if we adjust the boundary potential in such a way
that\ \eqref{askas}\ can be written in the form
\bea\label{askasii} \big[\, w_3-{\bar w}_3\big]_{\sigma=0}=\ri\
{\rd \theta_2\over \rd \tau}\ , \eea with some  boundary field
$\theta_2=\theta_2(\tau)$, then the charge \bea\label{alksl}
q_2=\int_{0}^\infty\rd\sigma\, \big(\, w_3(\tau,\sigma)+{\bar
w}_3(\tau,\sigma)\, \big)-\theta_2(\tau) \ , \eea will not depend
on $\tau$\ \cite{gz}: \bea\label{jasgj} {\rd q_2\over \rd\tau}=0\
. \eea This occurs  for $f\sim \re^{-2x}={1\over \cos^2(y)}$\,,
i.e., for the boundary potential such that \bea \label{ashskh}
U\to -{nC\over R}\ \tan(y)\ \ \ \ \ \ {\rm as}\ \ \ n\to\infty\ ,
\eea where $C$ is an arbitrary dimensionless constant.

If one takes   $\tau$  to be  the Euclidean time,
Eq.\eqref{jasgj}\ signifies a presence of nontrivial conservation
charge in the perturbed theory. As a matter of fact  the classical
hairpin  model with the boundary potential\ \eqref{ashskh}\
possesses an infinite number of conserving charges $q_s$ with
$s=1,\, 2,\, 3\ldots$\ . They are   generated by classical
holomorphic currents $w_{s+1}$ satisfying the condition
\bea\label{asaakas} \big[\, w_{s+1}- {\bar w}_{s+1}\,
\big]_{\sigma=0}= \ri\ {\rd \theta_s\over \rd \tau}\ . \eea For
$s=1$, \bea\label{alksjaiufdhew} w_2=-(\partial x)^2-(\partial
y)^2\ , \eea and the corresponding  charge $q_1$ coincides with
the energy. By means of  direct calculation it is not hard to find
an explicit form of $w_4$: \bea\label{alskasre}\nonumber
w_4&=&-(\partial x)^4- 5\, (\partial y)^4-6\, (\partial
x)^2(\partial y)^2-4\, \partial^2x\,
(\partial y)^2+4\, \partial^2y\,\partial x\partial y+\\
&&\partial^3 x\,\partial x+\partial^3 y\,\partial y\ .
\eea
It is useful to keep in  mind
simple ambiguities  in a
choice  of $w_{s+1}$ satisfying  Eq.\eqref{asaakas}.
First,  these currents
are defined up to  total
derivatives,
$w_{s+1}\to w_{s+1}+\partial g$ with ${\bar\partial} g=0$.
Second, the equation \eqref{asaakas}\ does not   fix an
overall multiplicative normalization of $w_{s+1}$.

\subsection{Local IM in the  AKNS soliton hierarchy}

Now we describe an   effective way to generate  all the densities
$w_{s+1}$ \eqref{asaakas} up to the above mentioned  ambiguities.
At this step we need to introduce the  fields \bea\label{aslsjs}
\psi=\ri\ (\partial y+\ri\, \partial x)\ \exp(2 \ri\, y_R)\, ,\ \
\ \psi^*=\ri\ (\partial y-\ri\, \partial x)\ \exp(-2\ri\, y_R)\ ,
\eea where  $y_R$ in the exponential stands for the holomorphic
part of the harmonic field $y$: \bea\label{klsklas}
y(\tau,\sigma)=y_R(\sigma+\ri\tau)+y_L(\sigma-\ri\tau)\ . \eea
Hence $\psi$ and $\psi^*$ are non locally expressed in terms of
the fundamental  field ${\bf x}$. They are clearly holomorphic
$({\bar
\partial}\psi={\bar \partial}\psi^*=0)$ and quasiperiodic fields:
\bea\label{hsjahgs} \psi(\tau+2\pi R)=\re^{-4\pi{\rm i} \xi_y}\
\psi(\tau )\, ,\ \ \ \ \psi^*(\tau+2\pi R)=\re^{4\pi{\rm i} \xi_y
}\ \psi^*(\tau)\ , \eea where $\xi_y$ is the second component of
the vector ${\boldsymbol\xi}$ in Eq.\eqref{saklskh}.

The fields\ \eqref{aslsjs} are
remarkable in many extents. First of all,
the  densities $w_s$ in
Eqs.\eqref{amsbj},\,\eqref{alksjaiufdhew},\,\eqref{alskasre}
are nicely expressed in terms of\ $\psi$ and $\psi^*$:
\bea\label{asjajs}
w_2&=&\psi\psi^*\,,\nonumber\\
w_3&=&{\textstyle{\ri \over 2 }}
\ \big(\psi^*\partial\psi-\psi\partial\psi^*\, \big)\ ,
\\ \nonumber
w_4&=&-{\textstyle{1\over 2}}\ \big(\, \psi^*\partial^2\psi+
\psi\partial^2\psi^*
\big) -(\psi\psi^*)^2\ .
\eea
More importantly, $\psi$ and $\psi^*$
generate  a closed  Poisson
subalgebra in the space of  classical fields.
To introduce the Hamiltonian picture here,
we  will interpret the world-sheet coordinate $\sigma$
as the Euclidean time. Then
the classical bulk action for the fundamental  field ${\bf x}$
defines a canonical Hamiltonian structure which
implies the following
set of  Poisson brackets for the holomorphic components:
\bea\label{ssass}\nonumber
\{\, x_R(\tau)\, ,\, x_R(\tau')\, \}&=&
\{\, y_R(\tau)\, ,\, y_R(\tau')\, \}=-{\textstyle{1\over 4}}
\ \epsilon(\tau-\tau')\, ,\\
\{\, x_R(\tau)\, ,\, y_R(\tau')\, \}&=&0\ ,
\eea
with
$$\epsilon(\tau)=2l+1\ \ \ {\rm for}\ \ \
2\pi R\,  l<\tau< 2\pi R\, (l+1)\, ;\ \ l\in{\mathbb Z}\ .
$$
Using Eqs.\eqref{ssass}\ it is easy to show that
\bea\label{akhkshk}
\nonumber
&&\{\, \psi(\tau)\, ,\, \psi(\tau')\,\}=
\epsilon(\tau-\tau')\ \psi(\tau)
\psi(\tau')\, ,\\
&&\{\, \psi^*(\tau)\, ,\, \psi^*(\tau')\,\}=\epsilon(\tau-\tau')\ \psi^*(\tau)
\psi^*(\tau')\ , \\ \nonumber
&&\{\, \psi^*(\tau)\, ,\, \psi(\tau')\,\}=\delta'(\tau-\tau')-
\epsilon(\tau-\tau')\ \psi^*(\tau)
\psi(\tau')\ .
\eea
The Poisson bracket algebra\ \eqref{akhkshk}\ is well known to describe
the second Hamiltonian structure for the
AKNS soliton hierarchy\ \cite{Magri,Faddeev}\ and
Eqs.\eqref{aslsjs}\ can be
interpreted as a transform to
the Dorboux variables\ \eqref{ssass} for this Hamiltonian structure.

The fields\ \eqref{asjajs}\ are local  densities of  the first
AKNS Hamiltonians: \bea\label{lsals} I_s^{\rm (class)}=\ri^{1-s}\
\int_0^{2\pi R}\rd\tau\ w_{s+1}(\tau)\ . \eea In particular
$I_2^{\rm (class)}$ coincides with the Hamiltonian of  the famous
non-linear Schr$ {\ddot {\rm o}}$dinger equation\ \cite{Zahar}\
\footnote{ The classical IM in Eq.\eqref{lsals} are normalized in
accordance with the convention from the book\ \cite{Faddeev}. If
$(x_R,y_R) $ in \eqref{aslsjs}\ are  real functions of the real
variable $\tau$, then $\psi$ and $\psi^*$ is a complex conjugated
pair and the corresponding  non-linear Schr$ {\ddot {\rm
o}}$dinger equation\ \eqref{kajdu} is in the repulsive regime. }.
There is an infinite number of  IM\ in the form\ \eqref{lsals}
which form  a commutative Poisson bracket subalgebra
\bea\label{ksahk} \{\, I_s^{\rm (class)}\, ,\, I_{s'}^{\rm
(class)}\, \}=0\ . \eea It is remarkable that all  AKNS   local
densities $w_s$\ \eqref{lsals} satisfy  Eq.\eqref{asaakas}. This
statement is essentially known in the literature in the context of
B${\ddot {\rm a}}$cklund transformation (see, e.g., Appendix B in\
\cite{Morosi}), even though it is not formulated in the language
of the  perturbed boundary theory.

It is well known (see, e.g., \cite{Faddeev})  that  the AKNS
flows, \bea\label{kajdu} {\partial\psi\over
\partial t_s}=\big\{\, I_s^{\rm (class)}\,,\, \psi(t_1,t_2\ldots)\,
\big\}\, ,
\ \ \ \ {\partial\psi^*\over
\partial t_s}=\big\{\, I_s^{\rm (class)}\,,\, \psi^*(t_1,t_2\ldots)\,
\big\}\ ,
\eea
describe isospectral deformations of the
first order differential operator
\bea\label{kjjskas}
{\cal L}=- \partial_{\tau}-{\textstyle {\rm i} \lambda\over  R}
\ H+\psi^*\ E+\psi\ F\ ,
\eea
where $\psi,\ \psi^*$ are  the quasiperiodic fields
on the segment $0\leq \tau\leq 2\pi R$\ \eqref{hsjahgs} and
$E,\, F$ and $H$ are the generators of the Lie algebra
$sl(2)$,
\bea\label{aasj}
[\, H\, ,\, E\,]=2\, E\, ,\ \ \
[\, H\, ,\, F\,]=-2\, F \, ,\ \ \
[\, E\, ,\, F\,]=H\ .
\eea
More precise, if we define the $(2j+1)\times (2j+1)$
monodromy matrices ${\bf M}_j(\lambda)$, $j=0,\, 1/2,\,
1,\,3/2\ldots$,
corresponding to the $(2j+1)$-dimensional representation $\pi_{j}$
of\ \eqref{aasj},  as
\bea\label{ksjdjsd}
\big({\boldsymbol
 \chi}_1(\tau+2\pi R)\, ,\ldots
{\boldsymbol \chi}_{2j+1}(\tau+2\pi R)\, \big)
=\big({\boldsymbol \chi}_1(\tau)\, ,\ldots
 {\boldsymbol \chi}_{2j+1}(\tau)\, \big)\
{\bf M}_j(\lambda)\ ,
\eea
where ${\boldsymbol \chi}_k(\tau)$
are
$(2j+1)$ linear independent solutions to the
auxiliary linear problem
\bea\label{alsksl}
\pi_{j}\big[\,{\cal L}\,\big]\, {\boldsymbol \chi}=0
\ ,
\eea
then the transfer-matrixes
\bea\label{lasks}
 T_j(\lambda)
={\rm Tr}_{\pi_j}\big[\, \re^{-2\pi {\rm i}
\xi_y H}\, {\bf M}_j(\lambda)\, \big]
\eea
are involutive (with respect the Poison
structure\ \eqref{akhkshk}) IM of the AKNS flows:
\bea\label{assks}
\{\,  T_j(\lambda)\, ,\, { I}_s^{\rm (class)}\, \}=0
\eea
and
\bea
\{\,  T_j(\lambda)\, ,\,  T_{j'}(\mu)\, \}=0\ .
\eea
As the matter of fact the transfer-matrices $ T_j(\lambda)$
are not independent for different $j$ and
can be algebraically  expressed
in terms of $ T(\lambda)\equiv T_{1\over 2}
(\lambda)$ corresponding to the fundamental representation of $sl(2)$.
The latter can be thought of as a generating
function of the local IM \eqref{lsals} as it expands in the
$\lambda\to\infty$ asymptotic series\ \cite{Faddeev}:
\bea\label{kjsjs}
 T(\lambda)=2\ \cosh\big(2\pi\nu(\lambda)\big)\ ,
\eea
with
\bea\label{laskjl}
 \ri\ \nu(\lambda)\simeq
-\lambda-\xi_y+{\textstyle{1\over 2\pi}}\
\sum_{s=1}^{\infty} I_{s}^{\rm (class)}\ \ \Big({R\over 2\lambda}\Big)^{s}\ \ \ \ \ \
{\rm as}\ \ \ \ \ \ \lambda\to\infty\ .
\eea

\section{Semiclassical quantization}
\label{sectwo}

Here we study  the semiclassical behavior   of the boundary
amplitude\ \eqref{lksjsa} using the path integral approach. For
the sake of discussion it is  convenient to consider the conformal
map of the semi-infinite cylinder, $\tau\equiv\tau+2\pi R,
\sigma\geq 0$, to the disk of radius $R$: \bea\label{sksjsk}
{z\over R}=\re^{-(\sigma+{\rm i}\tau)/R}\, ,\ \ \ \ {{\bar z}\over
R}=\re^{-(\sigma-{\rm i}\tau)/R} \eea
Then the overlap ${}_{\rm pert}\langle\,
 B\,  |\, {\bf P}\, \rangle$ with the Fock vacuum $|\,
{\bf P}\, \rangle$ relates to the unnormalized one-point function
of associated primary field inserted at the center of the disk,
\bea \label{exponent} \big\langle\,  \re^{\ri {\bf P}\cdot {\bf
X}} (0,0) \, \big\rangle_{\rm disk} =  R^{1/3-{{\bf P}^2/ 2} }\
{}_{\rm pert}\langle\, B\, |\, {\bf  P}\,  \rangle\,, \eea
and
$R^{1/3}\ {}_{\rm pert}\langle\, B\, |\, {\bf  0}\,  \rangle$ is
the disk partition function\footnote{We always assume that the
exponential field in \eqref{exponent} is defined according to the
usual CFT conventions \cite{Book}. The factor $R^{1/ 3}$ appears
due to the conformal anomaly.}. The one-point function\
\eqref{exponent} can be represented in terms of the functional
integral as follows, \bea\label{path} \big\langle\, \re^{\ri {\bf
P}\cdot {\bf X}} (0,0) \, \big\rangle_{\rm disk}=\int\, {\cal
D}X\,{\cal D}Y\ \re^{\ri PX + \ri QY}(0,0)\ \re^{-{\mathscr
A}[{\bf X}]- {\mathscr A}_{\rm pert}[{\bf X}_B]}\ , \eea where
${\bf P}=(P,Q)$, and the integration variables $X(z,{\bar
z}),Y(z,{\bar z})$ are assumed to obey the constraint\
\eqref{bconstaint} at the boundary $|z|=R$. Notice that at the
classical  limit the effect of  exponential insertion in\
\eqref{path} can be accounted for    by imposing the asymptotic
condition\ \eqref{saklskh} in the cylindrical frame. For this
reason the vector ${\boldsymbol\xi}$ in\ \eqref{saklskh}\ is
proportional to ${\bf P}$: \bea\label{saja} {\boldsymbol
\xi}={{\bf P}\over 2\sqrt{n} }\ . \eea When $P$ and $Q$ are pure
imaginary, some insight can be gained by making a shift of
integration variables, \bea\label{alksasa} {\bf X}\to {\bf
X}+\ri\, {\bf P}\ \log{|z|\over R}\ , \eea in the functional
integral\ \eqref{path}, which brings it to the form \bea
\label{exponentaa} \big\langle\,  \re^{\ri {\bf P}\cdot {\bf X}}
(0,0) \, \big\rangle_{\rm disk}=R^{-{\bf P}^2/2}\ \int\, {\cal
D}X\,{\cal D}Y\ \ \re^{-{\mathscr A}_{\rm bulk} [{\bf
X}]-{\mathscr A}_{\rm bound}[{\bf X}_B]}\ , \eea where the
boundary action is given by \bea \label{bounda} {\mathscr A}_{\rm
bound}=-\oint_{|z|=R}{\rd z\over 2\pi z}\ \big(\, PX_B+QY_B+ \ri
R\ U(X_B,Y_B)\, \big)(z)\ . \eea In this section we evaluate
leading semiclassical contribution to the overlap\ \eqref{lksjsa}\
by direct calculation of the functional integral\
\eqref{exponentaa} in  the saddle-point approximation. This will
give some intuition about its structure.

\subsection{``Light'' vertex insertion}

Making  an attempt to  calculate\ \eqref{path}
we   immediately   encounter  a problem;
The potential
\ \eqref{ashskh} is unbounded as $y\to\pm {\pi\over 2}$.
In fact, in order to ensure   a convergence of the functional integral
we must  place   the overall  pure imaginary constant $C$  in
\eqref{ashskh}\footnote{It deserves to draw reader's attention
that in \eqref{sskjd}
we use the same notation $\lambda$ as
for the spectral parameter in Eq.\eqref{kjjskas}.
The reason is  discussed at the
end of this section.}:
\bea\label{sskjd}
C=\ri \lambda\ .
\eea
For real positive $\lambda$,
the path integral\ \eqref{exponentaa} seems to be
well defined,
but the pure imaginary boundary potential $U({\bf X}_B)$
will certainly
break down the unitarity
of the underline QFT in the Minkowski coordinates $(\sigma,\,\ri\,\tau)$.

Let  us  assume that  the parameters $P,\, Q$ and $\lambda$
are
sufficiently small, i.e., the boundary action\ \eqref{bounda}
has no  appreciable effect on the saddle-point configuration.
To be  precise we write
\bea\label{kaksjkaj}
(P,\, Q)={\textstyle{2\over \sqrt{n}}}\ (p,\, q)
\eea
and assume that
\bea\label{alksjsa}
p,\, q\ \ \ {\rm and}\ \ \  n\lambda\sim 1\ \ \ \ {\rm as}\ \ \
n\to\infty\ .
\eea
Notice that these conditions are somewhat different
then those  considered in the previous section, where
the limit $n\to\infty$
was taken under the assumption that
$\xi_x=p/n,\, \xi_y=q/n$ and $\lambda$ are  fixed.

With the condition \eqref{alksjsa} the action is minimized by the
trivial classical solutions ${\bf X}(z,{\bar z}) = {\bf X}_0$,
where ${\bf X}_0 = (X_0 , Y_0)$ is an arbitrary point on the
hairpin\ \eqref{bconstaint}. Therefore \bea\label{mini} Z_{\rm
class} = \int_{\rm hairpin}\rd{\cal M}({\bf X}_0)\ \re^{2{\rm i}
(p{X_0} + {q Y_0})/\sqrt{n}}\ \re^{- R U({\bf X_0})}\ , \eea where
the integration measure $\rd{\cal M}({\bf X}_0)$ is determined by
integrating out  fluctuations around the classical solution in the
Gaussian approximation. Of course there is no need of actually
evaluating this Gaussian functional integral to figure out the
answer. If one writes ${\bf X}(z,{\bar z}) = {\bf X}_0 + \delta
{\bf X}(z,{\bar z})$, and splits the fluctuational part $\delta
{\bf X}$ into the components normal and tangent to the hairpin at
the point ${\bf X}_0$, the components are to satisfy the Dirichlet
and the Neumann boundary conditions, respectively. Therefore one
just has to take the product of the known (see, e.g., Appendix to
\cite{saleur}) disk partition functions with these two boundary
conditions. As
the result, $\rd {\cal M}({\bf X}_0)= {\textstyle{{\rm g}_D^2\over
2\pi}}\, \rd\ell({\bf X}_0)$, where $\rd\ell({\bf X}_0)
=\sqrt{(\rd X_0)^2 + (\rd Y_0)^2}$ is the length measure of the
hairpin, and ${\rm g}_D = 2^{-{1\over 4}}$ is  the ``boundary
degeneracy'' \cite{affleck} associated with the Dirichlet
conformal boundary condition\footnote{The definition is as
follows: ${\rm g}_D = \langle\,  B_D\, |\, {\bf P}\, \rangle$, where
$|\,B_D\,\rangle$ is the boundary state of {\it uncompactified}
boson $X$ with the Dirichlet boundary condition $X_B=0$, and the
primary states $|\, P\,  \rangle$ are delta-normalized,
$\langle\,P\,\mid \,P'\,\rangle = \delta(P -P')$.}.

Certainly the above analysis ignores the presence of (unbounded)
boundary potential. Its proper   account leads to the one loop
renormalization of the coupling constant $\lambda$. Indeed, the
form of the boundary potential in the classical limit  suggests
that
\bea\label{kjahs} U({\bf X}_B)\to \pm \ri\ {n\lambda\over  R}
\ \re^{-{X_B\over \sqrt{n}}}\ \ \ \ {\rm as}\ \ \ X_B\to-\infty\ ,
\eea
where the sign factor is  dictated by the choice of hairpin
asymptote $Y_B=\mp {\pi\over 2\sqrt{n+2}}$. The  anomalous
dimension of the boundary operator $\re^{-{X_B\over\sqrt{n}}}$
with respect to the hairpin stress-energy tensor \eqref{energymom}
is given by \bea\label{ehgsf} d_{\rm pert}=-{\textstyle {2\over n}}\ .
\eea If
one imposes now the  normalization condition
\bea\label{majhs} \Big[\, {\partial \over
\partial  E_*}\ \langle\, X_B\, \rangle_{\rm disk}\, \Big]_{E_*=0}=0\
\eea on the expectation value of  renormalized boundary field
$X_B$, then the renormalized coupling $\lambda$ is related to
the RG invariant scale $E_*$ as in Eq.\eqref{kajshk}.

Although
Eq.\eqref{kajshk} is a  result of the
semiclassical consideration, in the next section we shall bring
forward arguments which show that it is
perturbatively exact, i.e., a renormalization scheme exists in
which it is  exact to all orders in the ${1\over n}$-expansion.

All the above  add up   to the following integral representation
for the semiclassical boundary amplitude: \bea\label{kasjsl}
Z_{\rm class} ={\rm g}_D^2\ \sqrt{n}\ \big({\textstyle{n\over
2}}\, r^2\lambda\big)^{-{\rm i} p}\ \int_{-{\pi\over 2}}^{\pi\over
2} {\rd y\over 2\pi}\ \big(\cos(y)\big)^{2{\rm i} p-1} \,
\re^{2{\rm i} q \, y}\ \re^{{\rm i} n\lambda \tan (y)}\ , \eea
where   the additional factor $({\textstyle{n\over 2}}\,
r^2\lambda)^{-{\rm i} p}$ containing an arbitrary constant $r$ is
introduced to ensure the normalization condition\ \eqref{majhs}.
This factor can be also understood as an artifact of an additive
counterterm for the bare boundary field $X_B$. The integral\
\eqref{kasjsl}\ is expressed (see Eq.13, Section 6.11.2 in Ref.\cite{Bateman})
in terms of  Kummer's solution
$U(a,b;z)$ of the confluent hypergeometric equation\
\cite{Stegun}: \bea\label{nabsajash} Z_{\rm class}={\rm g}^2_D\
r^{-2{\rm i} p}\ \ {\sqrt{n}\ (2 n\lambda)^{-{\rm i} p} \over
\Gamma({1\over 2}-q+\ri p)}\ \ \ \re^{-n\lambda} \
U\big({\textstyle{1\over 2}}-\ri p+q,\, 1-2\ri p;\ 2n\lambda\,
\big)\ . \eea Here we assume that $\lambda>0$ and choose the
principal brunch of the multivalued function $U(a,b;z)$ such that
for  positive $z$ \bea\label{lasjs} U(a,b;z)\to z^{-a}\  \ \ \
{\rm as}\  \ \ \ z\to+\infty\ . \eea It is also instructive to
rewrite the semiclassical boundary amplitude \eqref{nabsajash} in
terms of the hypergeometric series  $M(a,b,z)=1+{az\over b}+{a(a+1)\over 2!
b(b+1)}\, z^2+ \ldots$\ : \bea\label{alasjsl} Z_{\rm class}=
Z_{\rm class}^{(-)}(p,q)\ F_{\rm class}^{(-)}(p,q\,|\, \lambda)+
Z_{\rm class}^{(+)}(p,q)\  F_{\rm class}^{(+)}(p,q\, |\, \lambda)\
, \eea
where \bea\label{lksjw} Z_{\rm class}^{(-)}(p,q)&=& {\rm
g}^2_D\, r^{-2{\rm i} p}\ \ \ {\sqrt{n}\ (2 n \lambda)^{-{\rm i} p}\
\Gamma(2\ri p)
\over \Gamma({1\over 2}+q+\ri p) \Gamma({1\over 2}-q+\ri p)}\, ,\\
Z_{\rm class}^{(+)}(p,q)&=&
{\rm g}^2_D\, r^{-2{\rm i} p}\ \ {\sqrt{n}\over\pi}\
(  2n\lambda)^{{\rm i} p}\
\cosh\big(\pi (p+\ri q)\big)\ \Gamma(-2\ri p)
\, ,\nonumber
\eea
and
\bea\label{alskj}
F_{\rm class}^{(\mp)}(p,q\, |\,\lambda)=
M\big({\textstyle{1\over 2}}\mp \ri p+q,\, 1\mp 2\ri p;\ 2n\lambda\big)\ .
\eea
In this form the nature of singular behavior as $\lambda\to 0$ is more
explicit. Let us make a (trivial) observation that the poles of
$Z^{(-)}_{\rm class}(p,q)$ in the variable $p$ in the first term in\
\eqref{alasjsl}
are canceled by the poles in the higher terms of the
confluent hypergeometric
series $ F^{(+)}_{\rm class}$ in the second term, and vice
versa, it makes  the full partition
function an entire function of parameters $p$ and $q$.

\subsection{``Heavy'' vertex insertion}

Although the above result \eqref{alasjsl} was derived under the assumption that $P$,$Q$
and $\lambda$
are small, it needs little fixing to become valid for much
larger values
of these parameters. When $(P,Q)=2\sqrt{n}\ (\xi_x,\xi_y)$
 become as large as
$\sqrt{n}$, \bea\label{asakjassa} \xi_x,\ \xi_y\ \  \ {\rm and} \
\lambda\ \sim 1\ \ \ \ {\rm as}\ \ \ n\to\infty\ , \eea the vertex
insertion in\ \eqref{path}\ and the boundary potential  is
``heavy'', i.e., it must be treated as a part of the action, and
they  affect the saddle-point analysis. The saddle-point
configuration(s) is still a constant field, but now ${\bf X}_0$ is
not an arbitrary point on the curve\ \eqref{bconstaint}, but has
to extremize the boundary action\ \eqref{bounda} \bea
\label{boundar} {\mathscr A}_{\rm bound}[{\bf X}_0]=-\ri \, {\bf
P}\cdot {\bf X}_0+ R\ U({\bf X}_0)\ . \eea There are two solutions
of the saddle-point equation which are  real valued for real
values of the parameter \bea\label{kajhsk}
\nu_0=\sqrt{\xi_x^2-2\lambda\,\xi_y-\lambda^2}\ . \eea One of them
corresponds to the minimum, and another to the maximum of $\ri\,
{\mathscr A}_{\rm bound}[{\bf X}_0]$. The saddle-point action
 produces nothing else but the $p,\, q\to\infty$ asymptotic
form of the expression\ \eqref{alasjsl} -- after all, this
asymptotic of the integral\ \eqref{kasjsl} is controlled by the
same saddle points. One can observe that if one splits the
constant-mode integration into two parts, as was suggested in\
\eqref{alasjsl}, the parts receive contributions from different
saddle points -- one from the ``minimum'' and one from the
``maximum'' one. What makes the difference at $(P,Q) \sim
\sqrt{n}$ is the proper treatment of non-constant modes. One
writes \bea\label{ksxusiu} {\bf X}(z,{\bar z}) = {\bf X}_* + {\bf
t}_* \ \delta X_t (z,{\bar z}) + {\bf n}_*\ \delta X_n (z,{\bar
z})\,, \eea where ${\bf X}_*$ is the position of the saddle point
on the hairpin, and ${\bf t}_*$ and ${\bf n}_*$ are unit vectors
tangent and normal to the hairpin at this point. Then for small
$\delta X_t$ the boundary constraint\ \eqref{bconstaint}\  reads
\bea\label{deltaxn} \delta X_n =-\re^{X_0\over \sqrt{n}}\ \ {
\delta X_{t}^2\over  2\sqrt{n}} + O(\delta X_{t}^3)\ , \eea and,
up to  higher-order terms, the boundary action \eqref{bounda} can
be written as \bea\label{bactionb} {\cal A}_B= A_{B}[\, {\bf
X}_*\, ] \mp\nu_0\ \oint \,{{\rd z}\over{2\pi z}}\ \delta X_{t}^2
\ , \eea with the coefficient $\nu_0>0$ is given by\
\eqref{kajhsk}; then the sign minus (plus) in \eqref{bactionb}
applies to the ``minimum'' (``maximum'') saddle point. Thus, while
to the leading approximation the normal component $\delta X_n$
still can be treated with the Dirichlet boundary condition, the
``boundary mass'' term in\ \eqref{bactionb}\ has to be taken into
account in evaluating the contribution from $\delta X_t$. Using
the well known boundary amplitude of the free field with quadratic
boundary interaction\ \cite{Witten}, one finds that
Eq.\eqref{alasjsl} would apply to the case of $(P,Q)\sim \sqrt{n}$
as well if one puts corresponding additional factors in the two
terms in\ \eqref{alasjsl}, i.e., replaces  $Z^{(\mp)}_{\rm
class}(p,q)$  there by \bea\label{shsdy} {\tilde Z}^{(\mp)}_{\rm
class}(P,Q) = {Z}^{(\mp)}_{\rm class}\big(\textstyle{\sqrt{n}
\over 2}\, P\, , \textstyle{\sqrt{n}\over 2}\, Q\big)\,
\Gamma\Big(1 \pm \ri\, \sqrt{{P^2\over n}- {4Q\over \sqrt{n}}\
\lambda-4\lambda^2}\ \Big) ; \eea of course in this case one can
use the $p,\, q \to\infty$ asymptotic forms of the factors
\eqref{lksjw} and \eqref{alskj}. More explicitly,
\bea\label{alsjskas} Z_{\rm class}\sim {{\rm g}_D^2\over
\sqrt{2\pi}}\ \bigg(\ {\Gamma(1+2\ri\nu_0)\over \sqrt{2\ri\nu_0}}\
\re^{{\rm i} n\, S_-}+ {\Gamma(1-2\ri\nu_0)\over
\sqrt{-2\ri\nu_0}}\ \re^{{\rm i} n\, S_+}\, \bigg)\ , \eea with
\bea\label{lkddii} \lambda\ {\partial S_{\pm}\over \partial
\lambda} =\pm  \nu_0\ . \eea Here the two terms in\
\eqref{alsjskas} correspond to the different saddle points of the
boundary potential\ \eqref{boundar}. The structure\
\eqref{alsjskas} is common for the semiclassical form  of Baxter's
$Q$-function (see Ref.\cite{Smir}). It is  instructive to note in
this connection that the  classical  transfer-matrix
$T^{\rm(vac)}(\lambda)$ corresponding to simplest (``vacuum'')
choice ${\bf x}_R={\ri {\boldsymbol \xi}\over R}\ (\sigma+\ri\tau)
$ in Eqs.\eqref{aslsjs},\,\eqref{kjjskas}, can be written in the
form \eqref{kjsjs}: $T^{\rm(vac)}(\lambda)=2\ \cosh(2\pi\nu_0)$,
where $\nu_0$ is the same function\ \eqref{kajhsk} as in
Eq.\eqref{lkddii}. One can easily recognize in  these equations
the classical counterpart of the Baxter relation\ \eqref{aksks}.

\section{Integrability of  the IPH model}

In this Section we would like to argue that the quantum IPH model
is integrable. Our first step toward the consistent quantum theory
is the quantization  of  the local IM of   AKNS soliton hierarchy.

\subsection{Quantum local IM}

Let us first   recall    structure  of the hairpin $W$-algebra\
\cite{LVZ}\,\footnote{ This   $W$-algebra was known for a long
while (see \cite{FATZ,FATT,Bakas}).}. Its generating currents $W_s
(u)\ (u=\sigma+\ri\tau)$ have spins $s=2,\, 3,\, 4\ldots$\ , and
can be characterized by the condition that they commute with two
``screening operators (charges)'', i.e., \bea\label{comscreen}
\oint_{u}\rd v \ W_s (u) \,{\cal V}_{\pm}(v) = 0\,, \eea where
\bea\label{screen} {\cal V}_{\pm} = \re^{\sqrt{n}\,X_R \pm
\ri\sqrt{n+2}\,Y_R} \eea and   $X_R$, $Y_R$ in the exponential
stand for the holomorphic part of the corresponding local fields.
The integration in Eq.\eqref{comscreen} is over a small contour
around the point $u$; vanishing of the integral\
\eqref{comscreen}\ implies that the singular part of operator
product expansion of $W_s (u)\,{\cal V}_{\pm}(v)$ is a total
derivative $\partial_v (\ldots )$. This condition fixes $W_s (u)$
uniquely up to normalization and adding derivatives and composites
of the lower-spin $W$-currents. For instance, the first
holomorphic current beyond\ \eqref{energymom}\ can be written as\
\eqref{wwwcurrent} where the ambiguity in adding a term
proportional $\partial W_2$ is fixed by demanding that\
\eqref{wwwcurrent}\ is a conformal primary. The higher currents
$W_4,\,  W_5,\ldots $ can be found either by a direct computation
of the operator product expansions with the screening
exponentials\ \eqref{screen}, or recursively, by studying the
singular parts of the operator product expansions of the lower
currents, starting with $W_3 (u) W_3 (v)$ and continuing upward.
Thus, the product $W_3 (u) W_3 (v)$ contains singular term $\sim
(u-v)^{-2}$ which involves, besides the derivatives $\partial^2
W_2$ and the composite operator $W_2^2$, the new current $W_4$.
Further operator products with $W_4$ define higher $W$'s, etc. In
this sense the $W$-algebra is generated by the two basic currents
$W_2$ and $W_3$.

As the matter of fact, there is the third, the most effective way to
generate all $W$-currents. It based on the observation
that the
screening operators associated with the exponentials
\eqref{screen}\ commute with
the parafermionic currents
\bea\label{parasha}
&\Psi (u)& =\ri\, \big(\, \sqrt{n+2}\ \partial Y  + \ri
\sqrt{n}\ \partial X \, \big)\
\re^{{{2{\rm i}}\over\sqrt{n+2}}\,Y_R(u)}\ , \\
\nonumber &\Psi^{*}(u)& =\ri\, \big(\, \sqrt{n+2}\ \partial Y  -
\ri\sqrt{n}\ \partial X \, \big) \re^{-{{2{\rm
i}}\over\sqrt{n+2}}\,Y_R(u)}\ , \eea in the same sense as the
$W$-currents do, i.e., \bea\label{psivertex} \oint_{u}\rd v\
\Psi(u)\,{\cal V}_{\pm}(v)  = 0\,,\ \  \qquad \oint_{u}\rd v\
\Psi^{*}(u)\,{\cal V}_{\pm}(v)  = 0\,. \eea The fields\
\eqref{parasha} extend the notion of the ${\mathbb Z}_k$
parafermions of \cite{ZamFat}\ to non-integer $k=-n-2$. Clearly
they can be also treated as a quantum version of the classical
nonlocal currents\ \eqref{aslsjs}. In spite the fact that the
fields\ \eqref{psivertex} are not local, both $\Psi$ and $\Psi^*$
are local with respect to the exponentials\ \eqref{screen}, hence
the integration contour in \eqref{psivertex} -- a small contour
around $u$ -- is indeed a closed one. It follows from
\eqref{psivertex} that all the fields generated by the operator
product expansion of $\Psi(u)\Psi^{*}(v)$ satisfy
Eqs.\eqref{comscreen}\ \cite{FATT,Bakas}. Thus we have
\bea\label{psiope} \Psi(u)\Psi^{*}(v) &=&  (u-v)^{-{2\over n+2}} \
\Big\{\, {\textstyle{n+2\over (u-v)^{2}} +{n\over 2}\
\big(W_2(u)+W_2(v)\big)} -\nonumber
\\ &&\textstyle{
{ \ri\, (u-v)
\over
2\sqrt{n+2}}\ \, \big(\, W_3(u)+W_3(v)\, \big)+\ldots}\, \Big\}\, ,
\eea
where $W_2$ and $W_3$ are the same as in\ \eqref{energymom}\ and
\eqref{wwwcurrent}, and the higher-order
terms involve the higher-spin $W$-currents.

After the brief  review
of the hairpin extended conformal
symmetry,
we turn to the description of the quantum AKNS local
integrals which  constitute
the Abelian subalgebra of the $W$-algebra.
The main idea is based on
examination of the asymptotic form of
classical boundary potential\ \eqref{kjahs};
It suggests to define the quantum local IM ${\mathbb I}_s$
as  elements of the hairpin $W$-algebra in the form\ \eqref{aks}
commuting with the additional  screening charge, i.e.,
\bea\label{kss}
 \oint_u \rd v\ P_{s+1}(u)\, {\cal V}_{0}(v)=\partial_u\big(\, \ldots\, )\ ,
\eea
where
\bea\label{salss}
{\cal V}_{0} = \re^{-{2X_R\over \sqrt{n}}}\ .
\eea

Remarkably  the condition\ \eqref{kss}\  indeed defines,
up to an overall multiplicative normalization, a local
integral of motion ${\mathbb I}_s$ for each $s=1,\, 2,\, 3\ldots$.
In particular, one can show that
$P_{2}=W_2$, $P_{3}=\ri\, W_3$
and
\bea\label{kash}
&&P_4=
n\, (\partial X)^4 +
6n\, (\partial X)^2 (\partial Y)^2 +
(5n+4)\,(\partial Y)^4+
\\ &&6 (n+1)\sqrt{n}\, \partial^2X (\partial Y)^2
(n^2+3n+1)\, (\partial^2 X)^2 +
(n^2+4n+2)\,(\partial^2 Y)^2 ,\nonumber
\eea
where we disregard all  terms which are total
derivatives and do not contribute to the ${\mathbb I}_3$.
Also it is   straightforward to check that
the such defined ${\mathbb I}_1,\, {\mathbb I}_2$
and ${\mathbb I}_3$
are mutually commute.
One may also note that in the classical limit
$P_4$  \eqref{kash}\
turns to be the
classical local density $(-w_4)$\ \eqref{alskasre}
up to a total derivative: $P_4\to n^3\
\big(\, -w_4+\partial(\ldots)\,
\big)$
as $n\to\infty$.

It seems likely that the formal proof of
existence and uniqueness solution of\ \eqref{kss} for $s=5,\, 6\ldots$  can be
obtained along the line of  Refs.\cite{jap,Eguchi,Frenk}.
Currently it is not available.
Nevertheless
we take the foregoing  facts as a strong
indication that an infinite sequence of commuting IM
$\{\, {\mathbb I}_s\, \}_{s=1}^{\infty}$ in the form
\eqref{aks}\ exists,
whose first local densities are given by
Eqs.\ \eqref{alsjs},\,\eqref{lkdpao} and \eqref{kash}.
It is expected  that the operators $ {\mathbb I}_s$
become the AKNS Hamiltonians\ \eqref{lsals}
in the classical limit.

\subsection{\label{secfour}
Diagonalization of  local IM}

It seems natural to take an attitude that the boundary state of
quantum  IPH model satisfies  the integrability condition\
\eqref{alksla}\ for all the local IM from  quantum AKNS series.
Thus we run into the problem
of simultaneous diagonalization of the AKNS local IM in the Fock
space ${\cal F}_{\bf P}$.

Let us consider  the normal mode expansion of the
quantum holomorphic current $\partial  {\bf X}$:
\bea\label{sakshs}
 \partial  {\bf X}={\textstyle {{\rm i} \over R}}\
\big(\,{\textstyle {1\over 2}}\, {\bf P}+
\sum_{k\not=0} {\bf X}_{k}\
\re^{k(\sigma+{\rm i}\tau)\over R}\ \big)\, .
\eea
Here ${\bf P}=(P,Q)$ is the zero-mode momentum
and $X^{\mu}_k=(X_k,Y_k)$ are the oscillatory  modes.
The canonical quantization procedure of the Poisson bracket algebra
\eqref{ssass}\ leads to the following set
of commutation relations,
\bea\label{alksu}
[\, X^{\mu}_k\, , X^{\nu}_s\, ]={\textstyle{k\over 2}}\ \delta^{\mu\nu}\ \delta_{k+s,0}\ .
\eea
In Appendix A we present explicit expressions
for the first local IM  in terms
of the oscillatory modes $X^{\mu}_k$.
As it follows from Eqs.\eqref{kjhdskjd},\,\eqref{basvs},
${\mathbb I}_s$ are well defined
operators acting in the Fock space  ${\cal F}_{\bf P}$.
The later is the highest weight module
over the Heisenberg algebra\ \eqref{alksu} with
the highest vector $|\, 0,\, {\bf P}\, \rangle$ (``vacuum'') defined
by the equation
\bea\label{shahksa}
 X^{\mu}_{k}\ |\, 0,\, {\bf P}\, \rangle=0\, ,\ \ \ \ \ k=1, 2, 3\ldots
\ .
\eea
The zero-mode momenta, $P$ and $Q$, acts in the given Fock space as $c$-numbers.
The space ${\cal F}_{\bf P}$ naturally splits into the
sum of finite dimensional ``level subspaces''
\bea\label{lksasl}
{\cal F}_{\bf P}=\oplus_{\ell=0}^{\infty}\,  {\cal F}^{(\ell)}_{\bf P}\ ;\ \
{\mathbb L}\, {\cal F}^{(\ell)}_{\bf P}=\ell\ {\cal F}^{(\ell)}_{\bf P}\ .
\eea
Since the grading operator ${\mathbb L}$ essentially coincides with ${\mathbb I}_1$,
\bea\label{kahsakjsa}
 {\mathbb I}_1=R^{-1}\ \big(\, {\mathbb L}+
{\textstyle {{\bf P}^2\over 4}}-
{\textstyle{1\over 12}}\, \big)\ ,
\eea
all the local IM act invariantly
in the level subspaces ${\cal F}^{(\ell)}_{\bf P}$. Therefore
diagonalization of $ {\mathbb I}_s$ in a given level subspace
reduces to a finite algebraic problem which
however rapidly becomes very complex
for higher levels. Notice that for $n\geq 0$ and real $P,\, Q$,
the operators ${\mathbb I}_s$ are hermitian
with respect to the canonical conjugation $(X^{\mu}_k)^\dagger=X^{\mu}_{-k}$,
so their spectra are real and their  eigenvectors are orthogonal to each other.
Here we list only the
vacuum eigenvalues for the first few $ {\mathbb I}_s$:
\bea\label{skkjashk}
{\mathbb I}_s\, |\, 0,\, {\bf P}\, \rangle=
R^{-s}\ I^{({\rm vac)}}_s(P,Q)\,  |\, 0,\, {\bf P}\, \rangle
\ ,
\eea
where
\bea\label{kajhs}
I_1^{({\rm vac)}}&=&{P^2\over 4}+{Q^2\over 4}-{1\over 12}\ , \nonumber\\
I_2^{({\rm vac)}}&=&{ Q\over 12}\ \big(\, (3 n+2)\, Q^2+3n\, P^2-
2n-1\, \big)\ ,\\
I_3^{({\rm vac)}}&=&{n\over 16}\ \big(\,
P^4-P^2+{\textstyle{1\over 12}}\, \big) +{3n\over 8} \,
\big(P^2-{\textstyle{1\over 6}} \big)\,
\big(Q^2-{\textstyle{1\over 6}}\big)+\nonumber\\ && (5n+4)\,
\big(\, Q^4-Q^2+{\textstyle{1\over 12}}\, \big)+
{{(n+3)(2n+1)\over 240}}\ .\nonumber \eea An explicit form of
the  eigenvalues and the corresponding eigenvectors in ${\cal
F}_{\bf P}^{(\ell)}$ is somewhat cumbersome even for small $\ell$.
For this reason we do not present it here. However, it suggests
that even the first three
local IM resolve all degeneracies in ${\cal F}_{\bf P}$, i.e.,
they orthonormalized eigenvectors  $\{|\, \alpha,\, {\bf P}\, \rangle\}$,
which  are labeled  by some  index
$\alpha$,
form  a basis  in the
Fock space.
This is in turn leads  to the structure \eqref{ssshsaksa}. Notice that
the amplitude  $B_0({\bf P})$ in Eq.\eqref{ssshsaksa}
corresponding to the normalized highest vector
$|\, 0,\, {\bf P}\,
\rangle\in{\cal F}_{\bf P}$,
\bea\label{aksajsh} \langle\,  {\bf P}',\, 0\, |\, 0,\,{\bf P}\,
\rangle=\delta({\bf P}'-{\bf P})\ , \eea
coincides with the complex conjugated overlap $Z^*$\ \eqref{lksjsa}.

\section{Dual form of  the IPH model}

Here we discuss an  alternative description of the IPH model and
explore the short distance behavior of the boundary amplitude\
\eqref{lksjsa} beyond the semiclassical approximation.

\subsection{Dual Hamiltonian}

In the recent work\ \cite{lzdual}  it was suggested that the hairpin model admits
equivalent ``dual''
description.
The dual model
involves a two-component Bose field  $\big(X(\sigma,\tau),{\tilde
  Y}(\sigma,\tau)\big)$ (where ${\tilde Y}$ is interpreted as the
T-dual of $Y$\,\footnote{The T-dual of  free massless field is defined as usual,
through the relations
$$Y=Y_{R}(\sigma+\ri\tau)+Y_{L}(\sigma-\ri\tau)\, ,\ \ \ \ \
{\tilde
Y}=Y_{R}(\sigma+\ri\tau)-Y_{L}(\sigma-\ri\tau)\ .$$
}) on the semi-infinite cylinder, which
has the free-field dynamics in the bulk, and obeys
{\it no constraint} at the boundary $\sigma=0$;
instead it interacts with an additional ``boundary'' degree of
freedom.
In the dual description it is convenient (but not necessary) to use
the Hamiltonian picture where  the cyclic coordinate $\tau\equiv
\tau+2\pi R$  is treated
as
the Matsubara (compact Euclidean) time $\tau$.
In this picture
the boundary amplitude $
Z^{(-)}(P,Q)= {}_{\rm hair}\langle\, B\, |\, {\bf P}\, \rangle$
admits the dual representation as the
trace
\bea\label{tracce}
Z^{(-)}(P,Q) = {\rm Tr}_{{\tilde{\cal H}}}\, \Big[\,
\re^{-2\pi R{\hat H}_{\rm hair}} \, \Big]
\,,
\eea
taken over the space ${\tilde{\cal H}} = {\cal
  H}_{X,{\tilde Y}}\otimes {\mathbb C}^2$, where ${\cal H}_{X,{\tilde
    Y}}$ is the space of states of the two-component boson
$\big(X(\sigma), {\tilde Y}(\sigma)\big)$ on the half-line $\sigma
\geq 0$ (with no constraint at $\sigma=0$) and ${\mathbb C}^2$
is the two-dimensional space representing the new boundary
degree of freedom.
 The dual Hamiltonian in \eqref{tracce} consists of
the bulk and the boundary parts,
\bea\label{alsajl}
{\hat H}_{\rm hair}&=&{\hat H}_{\rm bulk}-
{\textstyle{{\ri} \over 2\pi R}}\,  PX_{B}-
{\textstyle {{\ri} \sqrt{n+2}\, \over 4R}}\,  Q\sigma_3+
\\ &&\mu_-\, \Big[\,
\sigma_+\, \re^{{\sqrt n\over 2} X_B+{\rm i}{\sqrt {n+2}\over 2}\,
{\tilde Y}_B}+ \sigma_-\, \re^{{\sqrt n\over 2} X_B-{\rm i}{\sqrt
{n+2} \over 2}\, {\tilde Y}_B}\, \Big] \nonumber \ .\eea
The bulk part,
${\hat H}_{\rm bulk}$, is a ``free'' Hamiltonian corresponding to the
bulk action \eqref{baction}.  The
boundary term describes coupling of the boundary values $(X_B,
{\tilde Y}_B)\equiv
(X,{\tilde Y})|_{\sigma=0}$ of the field operators to the additional boundary
degree of freedom represented by ${\mathbb C}^2$ ($\sigma_{\pm}$ and
$\sigma_3$ are the Pauli matrices acting in ${\mathbb C}^2$).
We also include
in Eq.\eqref{alsajl}  an extra  parameter $\mu_->0$.
In the unperturbed hairpin model $\mu_-$ can be always eliminated by
shifting of the field $X$, but this coupling will be of
considerable importance for the IPH model.

We emphasize the   similarity of  the boundary vertex
operators
in Eq.\eqref{alsajl}
to  the screening charges \eqref{screen}  determining
the extended conformal symmetry of the hairpin model.
To construct the dual Hamiltonian for the perturbed
theory,
it is  crucial to observe
that the local IM  ${\mathbb I}_s$ defined through  the
condition\ \eqref{kss},\ also commute with the additional ``dual'' screening charge:
\bea\label{kssui}
 \oint_u \rd v\ P_{s+1}(u)\, {\cal V}^{(\rm dual)}_{0}(v)=\partial_u\big(\, \ldots\, )\ ,
\eea where \bea\label{salssaaxa} {\cal V}^{(\rm dual)}_{0} =
\re^{-2\sqrt{n} X_R}\ . \eea For $s=1,\, 2,\, 3$ Eq.\eqref{kssui}
was checked by the direct calculation using the explicit forms of
${\mathbb I}_s$. We take this as a strong indication that it holds
for all the local IM from  AKNS series. This in turn suggests a
possibility of dual description of the IPH model by means of the
Hamiltonian ${\hat H}_{\rm iph}\, :\ {\tilde {\cal H}}\to{\tilde {\cal H}}$,
  \bea\label{laaksjsa} {\hat H}_{\rm iph}= {\hat
H}_{\rm hair}+{ \Sigma}\ \re^{-\sqrt{n} X_B}\, ,
\eea
where
${\Sigma }$ is some $2\times 2$ matrix acting in
the ${\mathbb C}^2$-component of the Hilbert space ${\tilde {\cal H}}$.
In
what follows we will  argue that the matrix ${\Sigma}$
has the diagonal form, \bea\label{alksj} {
\Sigma }=\mu_+\re^{-{\pi{\rm i} n\over 2}\, \sigma_3}\ , \eea where
$\mu_+$ is an arbitrary constant. Notice that the Hamiltonian
${\hat H}_{\rm iph}$\ \eqref{laaksjsa}, \eqref{alksj} is not
hermitian. This is consistent with our previous observation from
Section\,\ref{sectwo} that the IPH model, once considered  in the Minkowski space,
is  a nonunitary QFT.

\subsection{Short distance expansion}

To justify the conjectured form of  dual Hamiltonian we shall
study the short distance behavior of the partition function corresponding to
\eqref{laaksjsa}:
\bea\label{tracceaa}
Z = {\rm Tr}_{{\tilde{\cal H}}}\, \Big[\,
\re^{-2\pi R{\hat H}_{\rm iph}} \, \Big]
\ .
\eea
Qualitatively, one may expect
that if $\Im m\, P$ is not too small, the limit $R\to 0$ of the
partition function $Z$ is controlled by either the hairpin
boundary operators\ $\re^{{\sqrt n\over 2} X_B\pm{\rm i} {\sqrt
n+2\over 2}Y_B}$ \eqref{alsajl} or the Liouville boundary vertex\
\eqref{laaksjsa} depending on the sign $\Im m P$, with some
crossover at small $\Im m\, P$.

More precisely, let us assume first that
$\Im m P<0$. Then we can treat the second term in\ \eqref{laaksjsa} as
a perturbation
and in the  leading approximation
\bea\label{kahssha}
Z|_{R\to 0}\to Z^{(-)}(P,Q)\ \ \ \ \ \  (\,\Im m\, P<0\,)\ ,
\eea
where $ Z^{(-)}(P,Q)={}_{\rm hair}\langle\, B\, |\, {\bf P}\, \rangle$  is  the
hairpin boundary amplitude \eqref{tracce}\ found in the paper \cite{LVZ}:
\bea\label{kjsahs}
Z^{(-)}=
 {\rm g}_D^2\ \Big(2\pi R\, {\mu_-^2\over  n}\Big)^{-  {2{\rm i} p\over n}
}
 \ {{\sqrt{n}\ \Gamma(2\ri\, p)\, \Gamma(1 +
{2{\rm i} p\over n})\over \Gamma\big({1\over 2} +q+  \ri\, p)\,
\Gamma({1\over 2} -q+  \ri\, p )}}\ . \eea Here and bellow we
often use the notations \bea\label{lkask} p={\textstyle{1\over
2}}\ \sqrt{n}\ P\, ,\ \ \ \ q={\textstyle{1\over 2}}\ \sqrt{n+2}\
Q\ . \eea Notice that in the semiclassical case $n\gg 1$ the
parameters $p,\, q$\ \eqref{lkask} are the same as in\
\eqref{kaksjkaj}; One should not distinguish between $n$ and $n+2$
at the perturbative order discussed in Section\,\ref{sectwo}.

Now we turns to the case $\Im m\, P>0$,
where the leading short distance behavior is
controlled by the Liouville vertex $\re^{-\sqrt{n} X_B}$.
In fact, because of the matrix factor ${\Sigma}$\ \eqref{alksj},
there are two noninteracting
boundary  Liouville field theories with the boundary ``cosmological constants''
$\mu_+\, \re^{-{\pi{\rm i} n\over 2}}$
and $\mu_+\, \re^{{\pi{\rm i} n\over 2}}$.
Using the result from the works\cite{FZZ,Tesh}
one has,
\bea\label{alkash}
Z|_{R\to 0}\to Z^{(+)}(P,Q)\ \ \ \ \ \  (\,\Im m\, P>0\,)\ ,
\eea
with
$$Z^{(+)}={\rm g}_D^2\, {\sqrt{n}\over 2\pi}\,
 \Gamma\big(-2\ri\, p \,)
\Gamma\big(1-2\ri\, {\textstyle{ p\over n}}\,
)\,
\Big({2\pi\mu_+ R^{n+1}\over \Gamma(1-n)}\Big)^{2{\rm i}  p\over n }\,
 \big(\, \re^{{\rm i} \pi q+\pi p}+\re^{-{\rm i} \pi q-\pi p}\, \big)\, .
$$
Here  the last factor  comes from
the trace over ${\mathbb C}^2$-component of the Hilbert space ${\tilde {\cal H}}$.

Since $Z^{(\pm)}(P,Q)$ vanishes in the limit $R\to 0$ if $P$ is
taken in the ``wrong'' half-plane (note the factors $R^{-{\rm i}\,
{ 2p\over n}}$ and $R^{2{\rm i} p\, {(n+1)\over n}}$ in
\eqref{kjsahs},\,\eqref{alkash}), this in turn suggests that the
overall $R\to 0$ asymptotic of the partition function is correctly
expressed by the sum \bea\label{jauy} Z|_{R\to 0} \to Z^{(-)}
(P,Q)+ Z^{(+)} ( P,Q)\, . \eea
Remarkably that Eq.\eqref{jauy} is
in a perfect agreement with our semiclassical result\
\eqref{alasjsl},\,\eqref{shsdy} provided \bea\label{kajhsktrs}
{\mu_-^2\over \mu_+}={n\over \Gamma(1-n)}\ \big(r^2 R\big)^{n}\ ,
\eea and \bea\label{lsjsa} \mu_-^2\mu_+={\Gamma(1-n)\over 16\pi^2
n }\  (2 E_*)^{n+2}\ , \eea where  $E_*$ is the RG invariant
``physical scale'' in the IPH model\ \eqref{kajshk}. It gives a
strong support to the conjectured Hamiltonian\ \eqref{laaksjsa}.

What can be said about corrections to the leading asymptotic?
Again we assume that $\Im m \, P<0$ and consider the perturbative
effect of the second  term in\ \eqref{laaksjsa} to the hairpin
boundary amplitude\  \eqref{kahssha}. In the unperturbed  hairpin
theory   the parameter $\mu_-$ is a dimensionless  constant. Let
us   eliminate $\mu_-$  from the hairpin Hamiltonian by   shifting
of the field $X$. Then the coupling $\mu_+$ in front of the
Liouville term in Eqs.\eqref{laaksjsa},\,\eqref{alksj} is replaced
by $\mu_-^2\mu_+$. Since the  anomalous dimension of the boundary
operator $\re^{-\sqrt{n}X_B}$ with respect to the hairpin
stress-energy tensor \eqref{energymom} is given by
\bea\label{ehgsfs} d_{\rm dual}=-n-1\ , \eea then
$\mu_-^2\mu_+\sim\ E_*^{n+2}$, which is in an agreement with
Eq.\eqref{lsjsa}. Hence we deduce that the perturbative
corrections to the leading asymptotic\ \eqref{kahssha}\ should be
in a form of power series expansion of the dimensionless parameter
$\kappa^{n+2}$, with \bea\label{salsj} \kappa=E_*R\ . \eea The
case $\Im m \, P>0$ can be analyzed similarly and one comes to the
same conclusion about the form of perturbative corrections to the
leading asymptotic\ \eqref{alkash}.

The power series expansion  in $\kappa^{n+2}$ becomes invisible in
the   limit $n\to\infty$, while  the consideration from
Section\,\ref{sectwo}\ suggests that the short distance expansion
should be in a form of power series of the dimensionless parameter
$\lambda\sim \kappa^{n+2\over n}$\ \eqref{kajshk}. This structure
is neatly captured by the form \bea\label{structure} Z =
Z^{(-)}(P,Q)\  F^{(-)}(\, P,Q\, |\, \kappa\, ) + Z^{(+)}(P,Q)\
F^{(+)}( P,Q\, |\, \kappa\, )\,, \eea where the functions $
F^{(\pm)} (\, P,Q\, |\, \kappa\, )$ are  double power series in
integer powers of $\kappa^{n+2}$ and $\kappa^{n+2\over n}$:
\bea\label{doubleser} F^{(\pm)}(\,P,Q\,|\,\kappa\,) =
\sum_{i,j=0}^{\infty}\ f^{(\pm)}_{i,j}(\,P,Q\,)\ \kappa^{i\,
(n+2)+j\, {(n+2)\over n}}\ , \eea with $f_{0,0} =1$.  The exact
splitting into two terms in\ \eqref{structure}\ is not easy to
justify on general grounds, but can be supported by the following
arguments.

First, recall that the such splitting in the semiclassical
expression\ \eqref{alasjsl}\ corresponds to isolating
contributions from two saddle points. More importantly, the full
expression has to take care of the poles of the factors
$Z^{(\pm)}(P,Q)$ -- one should  expect that the partition
function\ \eqref{structure} is an entire function of $P$. The
expression\ \eqref{structure}\  is the simplest form fit for this
job. The poles of $Z^{(+)}(P,Q)$ are located at the points in the
lower half of the $P$-plane where $\ri\, \sqrt{n}\,P$ or $\ri\,
\textstyle{P\over \sqrt{n}}$ take non-negative integer values. At
these points the factor $\textstyle{ \kappa^{\ri P {(n+2)\over
2\sqrt{n}}}}$ in $Z^{(+)}(P,Q)$ ``resonates'' with certain terms
of the expansion \eqref{doubleser} in the first  term in\
\eqref{structure}. Therefore, the poles of the factor
$Z^{(+)}(P,Q)$ in the second  term can (and must) be cancelled by
poles in appropriate terms of the expansion of the first term. For
the ``perturbative'' poles at $\ri\, \sqrt{n}\,P = 0,\, 1,\,
2\ldots $ this mechanism is evident in the semiclassical
expression \eqref{alasjsl}. The form similar to \
\eqref{structure}, together with this mechanism of the pole
cancellation,  was previously observed in the boundary sinh-Gordon
model \ \cite{AlZ} and in the paperclip theory \cite{LVZ}.

With the dual representation \eqref{laaksjsa}\,,\eqref{tracceaa}
it is possible to make some
quantitative predictions about the coefficients in  expansions
\eqref{doubleser}. In particular,
using the approach  outlined in  Ref.\cite{Goulian} (see also \cite{Zam,lzdual}),
it is possible to obtain  Coulomb gas integral representations
for  $f^{(-)}_{i,0}(P,Q)$ at the
poles $P=\ri\, { m\over \sqrt{n}},\ m=0,\, 1,\, 2\ldots$
of the first term in the sum\ \eqref{structure}. Unfortunately,
for general $m$ and $i$
the such integral formulas appear to be too cumbersome for
any  practical use.
The sole exception is the case with $m=0,\, i=1$:
\bea\label{sdkjshs}
f^{(-)}_{1,0}\big|_{P=0}=-{2^n\Gamma(1-n)\over  4\pi^2 n}\
{J(q)\over 2\cos(\pi q)}\ ,
\eea
where
\bea\label{lijsl}
J(q)&=&
 \re^{{\rm i}\pi (q-{ n\over 2})}\, J_1(q)+
\re^{-{\rm i}\pi( q-{ n\over 2})}\, J_1(-q)+
\re^{{\rm i}\pi (q+
{ n\over 2})}\, J_2(q)+\\
&&\re^{-{\rm i}\pi (q+{ n\over 2})}\, J_2(-q)+
\re^{{\rm i}\pi (q-{ n\over 2})}\, J_3(q)+
\re^{-{\rm i}\pi (q-{n\over 2})}\, J_3(-q)\nonumber
\eea
and
\bea
&&J_1=\int_{\atop
0<w<v<u<2\pi}
\Big[2\sin\big({\textstyle{u-v\over 2}}\big) \Big]^{-n-1}
\Big[4\sin\big({\textstyle{u-w\over 2}}\big)\,
\sin\big({\textstyle{v-w\over 2}}\big)  \Big]^{n}
 \re^{{\rm i} q (v-u)}\nonumber\\
&&J_2=\int_{\atop
0<v<w<u<2\pi}
\Big[2\sin\big({\textstyle{u-v\over 2}}\big) \Big]^{-n-1}
\Big[4\sin\big({\textstyle{u-w\over 2}}\big)\,
\sin\big({\textstyle{w-v\over 2}}\big)  \Big]^{n}
 \re^{{\rm i} q (v-u)}\nonumber\\
&&J_3=\int_{\atop
0<v<u<w<2\pi}
\Big[2\sin\big({\textstyle{u-v\over 2}}\big) \Big]^{-n-1}
\Big[4\sin\big({\textstyle{w-u\over 2}}\big)\,
\sin\big({\textstyle{w-v\over 2}}\big)  \Big]^{n}
 \re^{{\rm i} q (v-u)}\,.\nonumber
\eea
These integrals should be understood in a sense of  analytical continuation
in $n$ from the domain of convergence $(-1<\Re e\, n<0).$
The phases  in \eqref{lijsl}\ are also worthy of notice.
They are generated by the  additional boundary
degree of freedom  in\ \eqref{alsajl}
and make the sum of integrals\ \eqref{lijsl}\ well defined.
Were  these phases absent  $J(q)$ would depend
on an auxiliary initial point of integration
for the variables $u,\,v,\, w$. Here this
point is chosen to be zero.
For  the  given phase factors
the sum of integrals\ \eqref{lijsl} can be expressed
in terms of the generalized hypergeometric function at unity to yield
\bea\label{awelksj}
f^{(-)}_{1,0}\big|_{P=0}={2^n\Gamma^2(-n)\Gamma({1\over 2}+q)\over
\Gamma({1\over 2}+q-n)}\
 {}_3F_{2}\big({\textstyle{1\over 2}}+q,\, -n,\, -n\,;\,
1,\, {\textstyle{1\over 2}}+q-n\, ;\, 1 \big) .
\eea

\section{ The IPH model for $n\to 0$\label{secsix}}

The dual description of the IPH model is especially useful in the
strong coupling ($n\to 0$) regime. In particular for $n=0$ the
theory simplifies drastically and can be explored with  full
details.

\subsection{Boundary amplitude}

Here we consider the boundary amplitude\ \eqref{lksjsa}\ in the
$n\to 0$ limit, assuming the parameters $p,\,q$\ \eqref{lkask} are
fixed. For $n=0$ the field $X$ formally decouples in the dual
Hamiltonian\ \eqref{alsajl},\,\eqref{laaksjsa}.
To be  more precise the  $n\to 0$ limit
is, in fact, the classical limit for $X$. After the field
redefinition, \bea\label{lkdsl} \phi=\sqrt{n}\ X \eea this becomes
particularly striking. Since quantum fluctuations of  $\phi$ are
suppressed as $n\to 0$  we may  apply the saddle point
approximation to account  their  contribution. The theory is
trivial in the bulk and, in  consequence, the saddle point
achieves at  some  constant classical field configuration which we
shall denote as $\phi_0$.  Thus, in the limit $n\to 0$ we find
\bea\label{lakass} {\hat H}_{\rm iph}|_{n\to 0}&=& {\hat
H}_{\rm bulk}-{\ri p\over \pi n R}\ \phi_0+ \mu_+\re^{-\phi_0}+\\ &&h\,
\sigma_3+ \mu_- \re^{\phi_0\over 2}\ \Big[\, \sigma_+\, \re^{{\rm
i}\, {\sqrt {n+2}\over 2}\, {\tilde Y}_B}+ \sigma_-\, \re^{-{\rm
i}\, {{\sqrt {n+2}}\over 2}\, {\tilde Y}_B}\, \Big]_{n\to 0}\ ,
\nonumber \eea with \bea\label{fskjss} h= -{\textstyle {\ri  \over
2}}\ \big( {\textstyle {q\over R}}+\pi n\, \mu_+\,
\re^{-\phi_0}\big)\ . \eea One can observe that \eqref{lakass} is
a Hamiltonian of  the one-channel anisotropic Kondo model
\cite{Andersen} and $n=0$ corresponds to the  so called  Toulouse
point\ \cite{Toulouse}\footnote{ In the conventional   Kondo model
the field ${\tilde Y}$ emerges from the bosonization of   electron
degrees of freedom and is  assumed to be compactified.}. It is
well known that  the ground state energy of the  Kondo model
suffers from a specific ``free-fermion'' divergence at the
Toulouse point. For this reason we prefer to keep $n$ as an
ultraviolet regulator in Eq.\eqref{lakass} rather than set it to
zero. Notice also that  $h$\ \eqref{fskjss}\ plays the role of an
external magnetic field coupled with the impurity spin. The
partition function of  Kondo model is well known\ (see, e.g.,
Ref.\cite{Tsvelik}), \bea\label{salisus} Z_{\rm Kondo}={2\pi\ {\rm
g}_D\ \ \re^{-{2\pi R\over n}\, U_{\rm eff}(\phi_0)} \over
\Gamma({1\over 2}+2\ri R h+\pi R\mu_-^2\,  \re^{\phi_0}) \,
\Gamma({1\over 2}-2\ri R h+ \pi R\mu_-^2\,  \re^{\phi_0} )}\ ,
\eea where \bea\label{trelasks} U_{\rm eff}(\phi_0)=-{\ri p\over
\pi R}\ \phi_0+ n\mu_+\, \re^{-\phi_0}+{\textstyle{\mu_-^2\over
n}}\ \re^{\phi_0}+ n\ E_0\ . \eea The first two terms in the
effective potential $U_{\rm eff}$ came from the constant
($\phi_0$-depended) terms  in the Hamiltonian\ \eqref{lakass}. The
third term  is the above mentioned free-fermion divergence and
$E_0$ is a  non-universal constant with the dimension of energy.
The saddle point configuration $\phi_0$ corresponds to the minima
of the effective potential $U_{\rm eff}$, which picks at
\bea\label{jsahgs} \sinh(\phi_*)={\ri p\over \kappa}\ \ \ \ \ \ \
\ \ (\phi_0\equiv-n\log r+\phi_*)\ , \eea where we use the
notations $r$ and $\kappa$ from
Eqs.\eqref{kajhsktrs},\,\eqref{lsjsa},\,\eqref{salsj} with  $n\to0$.
Now we should  take into  account  an effect of  Gaussian
fluctuations around the classical saddle-point configuration. We can
write $\phi(\sigma,\tau)=\phi_0+\delta\phi(\sigma,\tau)$. Then
expanding the effective potential near the minima, we obtain the
boundary potential in the Gaussian approximation,
\bea\label{akjlashs} U_{\rm eff}(\phi_B)=const+{1\over 2\pi R}\
\sqrt{\kappa^2-p^2}\ \
(\delta\phi_B)^2+O\big((\delta\phi_B)^3\big)\ . \eea Finally we
can  use  the result from Ref.\cite{Witten} to find the
contribution of the  Gaussian field $\delta\phi$ into the
partition function. Combining all these ingredients together one
arrives to the following result \bea\label{aoksjo} Z|_{n\to
0}&=&{\rm g}^2_D\ r^{-2{\rm i} p} \ \
 {2\sqrt{\pi}\,
(\kappa^2-p^2)^{1\over 4}\
\Gamma(2\sqrt{\kappa^2-p^2})
\over \Gamma({1\over 2}-q+\ri p)
\Gamma({1\over 2}+q+\sqrt{\kappa^2-p^2})}\nonumber
\times \\ &&  \big( 2\kappa\re^{{\cal E}}\big)^{-\sqrt{\kappa^2-p^2}}\ \
\  \re^{-{S(\kappa,p)\over n}}\, ,
\eea
with
\bea\label{lksjl}
S=2\, \sqrt{\kappa^2-p^2}+2\,p\ \arcsin\big({\textstyle{p\over \kappa}
}\big)\  ,
\eea
and
${\cal E}$ is an arbitrary  non-universal constant.

\subsection{Boundary state  \label{secsix2}}

Not only the vacuum amplitude, but the whole boundary state (operator)
can be also found  in the limit
$n\to 0$. Contrary to
the previous discussion we shall keep
now the zero-mode momentum $P$ of order 1 for $n\to 0$.
Hence, in particular, one should send
$p=\sqrt{n} P/2\to 0$ in the expression \eqref{aoksjo} for
the vacuum amplitude.
With this
limiting prescription  $X$ and $Y$ sectors of the model
are factorized completely:
\bea\label{dvalkajs}
\lim_{n\to 0}{\mathbb B}=
 {\mathbb B}^{\rm (x)}\otimes  {\mathbb B}^{\rm (y)}\ .
\eea
Here ${\mathbb B}^{\rm (x)}$
is the boundary state operator corresponding
to the
Gaussian boundary interaction\ \eqref{akjlashs} with $p=0$,
and
${\mathbb B}^{\rm (y)}$ is the boundary state operator involving
the $Y$-modes only.
The operator ${\mathbb B}^{\rm (x)}$ admits a simple
representation in terms of the  oscillatory  modes $X_j$
\eqref{sakshs}\ \cite{sasha}:
\bea\label{kjshs}
{\mathbb B}^{\rm (x)}&=&{\rm g}_D\  r^{{{\rm i}P\over\sqrt{n}}}\ \
 \sqrt{\kappa\over\pi}\
\ \Gamma(2\kappa)\
\Big({2\kappa\over \re}\Big)^{-2\kappa}\
\ \re^{-{P^2\over 4\kappa}-\kappa{\cal E}_{\rm x}}\times\\ &&
\exp\Big(\, - \sum_{j=1}^{\infty}\ {\textstyle {4\over j}}\
 {\rm arctanh}\big(
{\textstyle{j\over 2\kappa}}\big)\ \ X_{-j}X_j\, \Big)\ ,\nonumber
\eea where ${\cal E}_{\rm x}$ is  some non-universal constant. The
boundary state operator ${\mathbb B}^{\rm (y)}$ does not have  such a
simple form    as \eqref{kjshs} for an arbitrary value of $q$.
Nevertheless its whole spectrum  is known.
As it was shown in Ref.\cite{blz2},  the
eigenvectors of ${\mathbb B}^{\rm (y)}$
 in the Fock space  of  representation
of the oscillatory  modes $Y_j$
 can be labeled by  two increasing
sequence of positive integers \bea\label{nsabvsj} 1\leq
n_1^{+}<n_2^{+}<\ldots <n_N^{+}\, ,\ \  1\leq
n_1^{-}<n_2^{-}<\ldots <n_N^{-}\ \ {\rm with} \  \ N\geq 0\, ,
\eea
and the corresponding eigenvalues are given by
\footnote{Eq.\eqref{jahsgj}
follows from the formula (4.28)  in Ref.\cite{blz2}, where
$2p_{BLZ}$ and $\lambda_{BLZ}$ should be replaced by the variables
$q+{\kappa\over 2}$ and $-\ri\ \sqrt{\kappa\over 2\pi}$
respectively.}: \bea\label{jahsgj} B^{(\rm y)}_{(n^+|n^-)}=B^{(\rm y)}_0\
&&\prod_{k=1}^{N} {(q+\kappa-n^{-}_k+{1\over 2})(q+{\kappa\over
2}+n_k^+-{1\over 2})\over
(q+\kappa+n^{+}_k-{1\over 2})(q+{\kappa\over 2}-n_k^-+{1\over 2})}\times\nonumber\\
&&\prod_{k=1}^{N}{(q+n^{+}_k-{1\over 2})(q+{\kappa\over 2}-n_k^-+{1\over 2})\over
(q-n^{-}_k+{1\over 2})(q+{\kappa\over 2}+n_k^+-{1\over 2})}\ ,
\eea
where
$$B^{(\rm y)}_0={\rm g}_D\ \
{2\pi
\over \Gamma({1\over 2}-q)
\Gamma({1\over 2}+q+\kappa)
}\ \Big({\kappa\over \re}\Big)^{\kappa}\ \re^{-\kappa({2\over n}+{\cal E}_{\rm y})}
\ ,
$$
and ${\cal E}_{\rm y}$ again is an arbitrary constant.

A limiting behavior of the boundary state operator
\eqref{dvalkajs} as $\kappa\to 0$ and
 $\kappa\to \infty$ deserves to be discussed in some details.
As it follows from Eq.\eqref{jahsgj},
\bea\label{ksaie}
\lim_{\kappa\to 0}{\mathbb B}^{(\rm y)}=
(\re^{{\rm i}\pi q}+\re^{-{\rm i}\pi q})\ \ {\rm g}_D\ {\mathbb I}\ ,
\eea
where ${\mathbb I}$ is the identity operator.
Clearly this  boundary state operator corresponds to the  Dirichlet
boundary condition for the  field $Y$ such that its boundary
values are constrained to two points,
\bea\label{kjsh}
Y_B|_{\rm UV}=\pm {\pi\over \sqrt{2}}\ .
\eea
This conclusion
is consistent with  a naive interpolation of the boundary constraint
\eqref{bconstaint}\ to $n=0$.
It is also clear that  the field $X$ is subjected by
the free boundary condition in the short distance limit:
\bea\label{kuisaie}
\lim_{\kappa\to 0}{\mathbb B}^{(\rm x)}={\rm g}_D\ \delta(P)\ \exp\Big(
2\ri\pi\ \sum_{j=1}^{\infty}{\textstyle {X_{-j}X_j\over j}}\, \Big)\ .
\eea

Now let us consider  the infrared behavior of the boundary state
operator\ \eqref{dvalkajs}. As $\kappa\to \infty$,
\bea\label{utewtksaie} {\mathbb B}^{(\rm x)}|_{\kappa\to \infty}\to
r^{ {{\rm i} P\over \sqrt{n}}}\ \ {\rm g}_D\ {\mathbb I}\  \
\re^{-\kappa{\cal E}_{\rm x}}. \eea This equation  immediately suggests
that the field $X$ obeys  the Dirichlet boundary condition in the
infrared limit: \bea\label{ldjdk} X_B|_{\rm
IR}={\textstyle{1\over \sqrt{n}}}\ \log r\ . \eea The infrared
boundary condition for the field $Y$ is far more interesting. As
it follows from Eq.\eqref{jahsgj}, \bea\label{sksajhs}
B^{(\rm y)}_{(n^+|n^-)}\big|_{\kappa\to\infty}\to {\rm g}_D\ \ \kappa^{-q}\ \ {\sqrt{2\pi}
\over \Gamma({1\over 2}-q)} \ \re^{-\kappa({2\over n}+{\cal
E}_{\rm y})}\ \ \prod_{k=1}^{N}{q+n^{+}_k-{1\over 2}\over
q-n^{-}_k+{1\over 2}}\ . \eea
If one looks at Eqs.\eqref{utewtksaie},\,\eqref{sksajhs}
closely,  several features stand out. First of all, it is expected
a priori that the infrared behavior of \eqref{dvalkajs} is
controlled by some  scale-invariant boundary state (operator). More
precise, the scale dependence of the infrared boundary state
(operator) is allowed in the form of a  non-universal factor
$\exp(-const\, \kappa)$\ with $\kappa=E_*R$. Such  factors are
indeed presented in Eqs.\eqref{utewtksaie},\,\eqref{sksajhs}. The
additional scale depended  factor $\kappa^{-q}$ in\
\eqref{sksajhs} shows that when the RG ``time'' $t=\log\kappa$
increases the field $Y$ ``flows'' uniformly with $t$. It can be
made into a  scale invariant fixed point by an appropriate
redefinition of the RG transformation, namely by supplementing it
with a formal field redefinition $(X,Y)\to (X,Y+{\ri\over
\sqrt{2}} \ {\delta t})$. This corresponds to the following
modification of  stress-energy tensor:
\bea\label{alksjo} T_{\rm
IR}&=&-(\partial X)^2-(\partial Y)^2 +{\ri\over \sqrt{2}}\
\partial^2 Y\, ,
\\
{\bar T}_{\rm IR}&=&-({\bar \partial} X)^2- ({\bar \partial} Y)^2
+{\ri\over \sqrt{2}}\ {\bar \partial}^2 Y\, ,\nonumber \eea or, in
a stringy speak, to introducing a complex linear  dilaton $D({\bf
X})= {\ri\over \sqrt{2}}\ Y$. In a view of  the non-unitary of
the IPH model in the Minkowski coordinates $(\sigma,\, \ri\,\tau)$
the  appearance of  complex dilaton is not particularly
surprising.

\section{Exact boundary amplitude}\label{secseven}

In this section we propose an exact expression for the boundary
amplitude\ \eqref{lksjsa}. The expression is in terms of solutions
of special second-order Ordinary Differential Equation (ODE).
Similar expressions are known in a number of integrable models of
boundary interaction, beginning with the work of Dorey and Tateo
\cite{toteo} (see, e.g.,\ \cite{LVZ} and references therein). Our
proposal extends this relation to the IPH model. In this case no
proof is yet available, but we will show in this section that the
proposed expression reproduces all the properties of the amplitude
described above.

\subsection{Differential equation}\label{secseven1}

Consider the ordinary second order differential equation
\bea\label{diff}
\bigg[\, -{{\rd^2}\over{\rd x^2}}   -
p^2+2q\, \kappa\, \re^{x}+\kappa^2\, \big(\re^{2x}+\re^{-nx}\big)
\, \bigg]\ \Psi(x) = 0\,,
\eea
where $p$, $q$ are related to the components of ${\bf P} =  (P,Q)$ in
\eqref{lksjsa} as in Eq.\eqref{lkask},
and $\kappa=E_{*}R$.
With this identification in mind, below we always assume that $\kappa$
is real and positive.

Let
$\Psi_{-}(x)$ be the solution of\ \eqref{diff}\ which decays when $x$ goes
to
$-\infty$ along the real axis, and $\Psi_{+}(x)$ be another solution
of\ \eqref{diff},
the one which decays at large positive $x$. We fix normalizations
of these two solutions as follows,
\bea
\label{psiassminus}
\Psi_{-} \to (2\kappa)^{-{1\over 2}}\
\exp\big(\, {\textstyle{nx\over 4}}-{\textstyle {2\kappa\over n}}\
\re^{-{nx\over 2}}\, \big) \qquad\qquad {\rm as} \qquad \quad x\to
-\infty\,,
\eea
and
\bea\label{psiassplus}
\Psi_{+}\to  (2\kappa)^{-q-{1\over 2}}\
  \exp\Big\{-\big(q+{\textstyle{1\over 2}}\big)\,x -
\kappa\, \re^{x}\, \Big\}
\ \,  \, \ \  {\rm as}\, \, \ \ \ x\to +\infty\, .
\eea
Let
\bea\label{wronskian}
{\rm W}[\Psi_{+},\Psi_{-}]\equiv \Psi_{+}\, {\rd\over \rd x}\,  \Psi_{-} -
\Psi_{-}\, {\rd\over \rd x}\,  \Psi_{+}
\eea
be the Wronskian of these two solutions. Then, our proposal
for the function\ \eqref{lksjsa}
is
\bea\label{exactz}
Z={\rm g}_D^2\ r^{-2{\rm i} p}
\ {\sqrt{2\pi}\over \Gamma({1\over 2}-q+\ri\, p)}\
\ {\rm W}[\Psi_{+},\Psi_{-}]\, .
\eea
To make more clear the motivations behind this proposal let us discuss
some
properties of the solutions of the differential equation\
\eqref{diff}\,\footnote{Eq.\eqref{diff}\ differs   in
the term $2q\kappa\, \re^{x}$ only  from the ODE
describing the boundary sinh-Gordon model (see Eq.(40) from the
work \cite{blz4}
with  $\beta^2<0$).
Because of this
the subsequent analysis is parallel  to the
one from the unpublished work Al.B. Zamolodchikov\ \cite{AlZ}.}.

\subsection{Small $\kappa$ expansion}\label{secseven2}

The ODE \eqref{diff} has
the form of one-dimensional Schr${\ddot{\rm o}}$dinger
equation with the potential  defined by the last
three  terms in the bracket in \eqref{diff}.
When $\kappa$ goes to zero this potential
develops a wide plateau at
\bea\label{plateau}
{\textstyle{2\over n}}\ \log\kappa\ll x\ll -\log\kappa\ ,
\eea
where its value is close to $-p^2$.
In this domain each of the solutions $\Psi_+$ and $\Psi_-$
is a combinations of two plane waves:
\bea\label{waves}
\Psi_{\pm}=D_{\pm}(p, q\, |\, \kappa)\ \re^{{\rm i} p x}+
D_{\pm}(-p, q\, |\, \kappa)\ \re^{-{\rm i} p x}\ .
\eea

Let us consider the amplitude $D_+$ in  \eqref{waves}\ for the
solution $\Psi_+$ first.
In this case it is convenient to change the variable,
$x=y-\log(2\kappa)$, and bring the equation \eqref{diff}
to the form
\bea\label{wekajshk}
\bigg[\, -{{\rd^2}\over{\rd y^2}}   -
p^2+q\,  \re^{y}+{\textstyle{1\over 4}}\, \re^{2y}+
\delta V_+\, \bigg]\ \Psi(x) = 0\,,
\eea
with
\bea\label{alkss}
\delta V_+=2^{n}\ \kappa^{n+2}\ \re^{-ny}\ .
\eea
Let us assume for  the moment that
\bea\label{lakhskp}
-2<\Re e\, n<0\ .
\eea
 Then the term
$\delta V_+$ can be considered as a small  perturbation for  any
$-\infty<y<+\infty$. At the zero perturbative order we   dismiss
$\delta V_+$ in \eqref{wekajshk}. It gives an equation which can
be brought to the form of   Kummer's equation\ \cite{Stegun} by a
simple change of  variables. The unperturbed solution reads
explicitly, \bea\label{lsj} \Psi^{(0)}_+(y)= \re^{-{\re^y\over
2}}\ \re^{-\ri p y}\ U\big({\textstyle{1\over 2}}+q-\ri p, 1-2\ri
p, \re^y\, \big)\ . \eea The normalization  of $\Psi_+^{(0)}$ is
chosen to match the asymptotic  condition\ \eqref{psiassplus}. Now
one can systematically develop the standard  perturbation theory
for the solution $\Psi_+$. Therefore   $D_+(p,q\, |\, \kappa)$\
\eqref{waves} admits the following power series expansion
\bea\label{skjs} D_+(p,q\, |\, \kappa)=(2\kappa)^{{\rm i} p}\ \ {
\Gamma(-2\ri p)\over \Gamma({1\over 2}+q-\ri p)}\
\Big(1+\sum_{i=1}^{\infty}d^{(+)}_i(p,q)\ \kappa^{i(n+2)}\, \Big)\
. \eea The first order perturbative coefficient $d^{(+)}_1$ can be
calculated in the closed form. Indeed, it is represented by the
following integral: \bea\label{lkslsl} d^{(+)}_1(p,q)&=&
{2^n\Gamma({1\over 2}+q-\ri p)\over \Gamma(1-2\ri p)}\
\int_0^{\infty}\rd z\
z^{-n-1}\ \re^{-z}\times\\
&& M\big({\textstyle{1\over 2}}+q-\ri p, 1-2\ri p, z\, \big)\
U\big({\textstyle{1\over 2}}+q+\ri p, 1+2\ri p, z\, \big)\
,\nonumber \eea which  converges for $-1<\Re e\, n<0$ and can be
expressed in terms of the generalized hypergeometric function at
unity (see Eq.(7.625) in Ref.\cite{Ryzhik}): \bea\label{sskahsk}
d_1^{(+)}(p,q)&=&{2^n\Gamma({\textstyle{1\over 2}}+ q-\ri
p)\Gamma(-2\ri p-n)\Gamma(-n)\over
\Gamma(1-2\ri p)\Gamma({\textstyle{1\over 2}}+q-\ri p-n)}\times\\
&& {}_3F_{2}\big({\textstyle{1\over 2}}+q-\ri p,\, -n,\, -n-2\ri
p\,;\, 1-2\ri p,\, {\textstyle{1\over 2}}-n+q-\ri p\, :\, 1\,
\big)\, .\nonumber \eea Evidently   the power series \eqref{skjs}\
converges for any complex $\kappa$ and defines an entire function
of $\kappa^{n+2}$
for  complex $n\not=-2+{1\over k}\
(k=1,\,2,\,3\ldots)$ from the strip\ \eqref{lakhskp}\,
\footnote{For $q=0$ the function $D_+(p,0\, |\,\kappa)$ coincides
up to the overall factor with a vacuum eigenvalue of the
$Q$-operator from the work \cite{blz2} upon  the parameter
identifications: $\beta^2_{BLZ}=-{n\over 2}$,\ $p_{BLZ}=-{{\rm i}
\over 2}\ p$ and $\lambda^2_{BLZ}=-{2^{-n-2} \over
\Gamma^2(1+{n\over 2})}\ \kappa^{n+2}$. In this connection, it is
pertinent to note that the higher-order coefficients in\
\eqref{skjs}\ can be  calculated by the method developed in
Appendix of the work\ \cite{blz5}.}. There is no reason to expect
the convergence for real positive $n$. However it is expected that
for $n>0$ Eq.\eqref{skjs}\ is valid in a sense of $\kappa\to 0$
asymptotic expansion with the zero convergence radius.

Similarly  one can derive  an expansion of the amplitude
$D_-$\ \eqref{waves}\ for the solution $\Psi_-$.
In this case we make a change of variable
$x={2\over n} \, z+{2\over n}\ \log({2\kappa\over n})$;
Eq.\eqref{diff} then takes the form,
\bea\label{ekajshk}
\bigg[\, -{{\rd^2}\over{\rd z^2}}   -
\big({\textstyle{2p\over n}}\big)^2+
 \re^{-2z}+
\delta V_-\, \bigg]\ \Psi(x) = 0\,, \eea with \bea\label{tyalkss}
\delta V_-={4q\over n}\ \Big({2\kappa\over n}\Big)^{n+2\over n}\ \
\re^{2 z\over n}+ \Big({2\kappa\over n}\Big)^{2(n+2)\over n}\ \
\re^{4 z\over n}\ . \eea For \bea\label{lakh} \Re e\, n<-2\ , \eea
$\delta V_-$ can be treated as a  perturbation for any real $z$.
The small parameter now is $\kappa^{n+2\over n}$. At the zero
order one has \bea\label{akshka} \Psi_{-}^{(0)}(z)=\sqrt{2\over
\pi n}\ K_{2{\rm i} p\over n}(\re^{-z})\ , \eea where $K_{\nu}(x)$
is the conventional MacDonald function. Again, the normalization
of the solution is chosen to match\ \eqref{psiassminus}. The
perturbative expansion for  $D_-(p,q\, |\, \kappa)$\
\eqref{waves}\ has a form: \bea\label{skjsu} D_-(p,q\, |\,
\kappa)=\Big({\kappa\over n}\Big)^{-{2{\rm i} p\over n}}\ \ {
\Gamma({2\ri p\over n})\over \sqrt{2\pi n}}\
\Big(1+\sum_{j=1}^{\infty}d^{(-)}_j(p,q)\ \kappa^{j\, {(n+2)\over
n}}\, \Big)\  . \eea This series  converges for any complex
$\kappa$  if $n$ belongs to the complex  half-plane\ \eqref{lakh}.
For real positive $n$ it is an asymptotic series with the zero
convergence radius. The low order coefficient  $d^{(-)}_1$ can be
easily obtained by direct perturbative calculations:
\bea\label{utykasjhsk} d^{(-)}_1(p,q)=2\, q\ \Big({2\over
n}\Big)^{2(n+1)\over n}\ {\Gamma({1\over 2}+{1\over
n})\Gamma(-{1\over n})\over 4\sqrt{\pi}}\ \ {\Gamma(-{1\over
n}+{2{\rm i} p\over n})\over \Gamma( 1+{1\over n}+{2{\rm i} p\over n})}\ .
\eea

The  asymptotic  expansions\ \eqref{skjs},\,\eqref{skjsu}
are useful in  study small  $\kappa$ behavior  of the
Wronskian\ \eqref{wronskian}. Indeed for small
$\kappa$ the Wronskian can be calculated
at any point from the domain\ \eqref{plateau},  where
$\Psi_{\pm}$ are  combinations of two plane waves\ \eqref{waves}.
Hence the Wronskian\ \eqref{wronskian}\ is
written in terms of the coefficients
in $D_{\pm}(p,q)$ as follows,
\bea\label{det}
&&{\rm W}\big[\Psi_{+},\Psi_{-}\big]=
 -2 \ri\,  p\times\\
&&\ \ \ \ \big(\, D_{+}(p,q\, |\, \kappa)\,D_{-}(-p,q\, |\,
\kappa) - D_{+}(-p,q\, |\,\kappa)\,D_{-}(p,q\, |\, \kappa)\,
\big)\, .\nonumber \eea Then\ \eqref{exactz}\ leads to the short
distance expansion  of  the partition function of the form \
\eqref{structure}, where $P,\, Q$ related to $p,\, q$ through
Eqs.\eqref{lkask}. The coefficient $f_{i,j}^{(\pm)}$\ in the
double power series \eqref{doubleser} are simply expressed in
terms of $d_{k}^{(\pm)}$: \bea\label{qqlksjslk}
f_{i,j}^{(\pm)}=d_i^{(+)}(\pm p,q)\ d_j^{(-)}(\mp p,q)\, , \eea
with $d_0^{(\pm)}=1$. In particular
$f_{1,0}^{(\pm)}(P,Q)|_{P=0}=d_{1}^{(+)}( 0, q)$, where
$d_{1}^{(+)}(p,q)$ is given by\ \eqref{sskahsk}. This is in a
complete  agreement with the result\ \eqref{awelksj}\ of
perturbative calculation in the IPH model.

\subsection{Semiclassical domain $\kappa\ll 1\ll n$}

When $\kappa$ is small  and  $n\gg 1$ the above expansions have to
be collected. This regime corresponds to the semiclassical domain
in the IPH model considered in Section\,\ref{sectwo}.

First, let us consider the case when $p$ and $q$ in \eqref{diff}
are order of 1;
it corresponds to the case of ``light'' insertion in\ \eqref{path}
with $(P,Q)\sim {1\over \sqrt{n}}$\ \eqref{alksjsa}. In this regime
the term $\kappa^2\,\re^{-nx}$ in the equation \eqref{diff}
has the effect of a rigid wall at some point
$x_0$, i.e., to the right from this point, for
$x-x_0\gg {1\over n}$, this term is negligible,
but to the left from $x_0$ it grows very fast, so
that for $x<x_0$ the solution $\Psi_-$ is essentially zero.
More precisely, when $x$ is above but  close to $x_0$ the
solution $\Psi_-$ is approximated by a
linear function,
\bea\label{sskshshs}
\Psi_-(x)\approx\alpha_0\ (x-x_0)\ ,
\eea
where the position of the wall $x_0$ and
the slope $\alpha_0$ are given by
\bea\label{lskjs}
x_0&=&{2\over n}\ \log\Big(
{\kappa\re^{\gamma_E}\over n}\Big)+o\big({\textstyle{1\over n}}\big)\ ,\\
\alpha_0&=&\sqrt{n\over 2\pi }\ \big(1+o(1)\, \big)\ ,\nonumber
\eea
as $n\to\infty$. Here $\gamma_E$ is  Euler's constant.
We will explain these relations shortly. Wronskian
\eqref{wronskian} is easily determined. Taking, for instance, any point
$x$ close to the right of the wall where both equations\ \eqref{lsj}\
and \eqref{sskshshs} are valid, one finds it equal $\alpha_0$ times the
expression \eqref{lsj}\ evaluated at $e^{y_0}=2\kappa\ \re^{x_0}$,
or, due to equation \eqref{lskjs}, at
$e^{y_0}=2n\lambda\big(1+o(1)\big)$,
where  $\lambda=({\kappa\over n})^{n+2\over n}$.
This leads exactly to\ \eqref{nabsajash}.

The case of ``heavy'' insertion, $(P,Q)\sim \sqrt{n}$\
\eqref{asakjassa},  can be handled similarly. The term
$\kappa^2\,\re^{-nx}$ still can be treated as a wall at $x_0$, in
the sense that at  $x-x_0\gg {1\over n}$ its effect is negligible,
but it dominates at $x_0-x\gg {1\over n}$. To study the vicinity of
the wall we make a change of the variable, $x={2\over n}\, z+
{2\over n}\ \log({2\kappa\over n})$, and bring the equation\
\eqref{diff} to the form\ \eqref{ekajshk}. With\ \eqref{ekajshk}
it is evident  that the wall located at $x_0={2\over n}\ \log(
{2\kappa\over n})+{2\over n}\, z_0+o({1\over n})$, where $z_0$ is
a some  $n$-independent constant. Close to the walls position,
i.e., at \bea\label{sjs} |x-x_0|\ll 1 \eea $\delta V_-$\
\eqref{tyalkss} can be approximated by the constant
\bea\label{sjljsl} \delta V_-\approx {8q\over n}\ \lambda+4\,
\lambda^2\ , \ \ \ \ \ {\rm where}\ \ \ \
\lambda=\Big({\kappa\over n}\Big)^{n+2\over n}\ . \eea Therefore
in the domain\ \eqref{sjs}\ the solution  $\Psi_-$ is approximated
by the MacDonald function, \bea\label{lksks} \Psi_-\approx
\sqrt{2\over\pi n}\ K_{ 2{\rm i}\nu_0}(\re^{-z})\ , \eea with
$\nu_0={1\over n}\ \sqrt{p^2-2q\, n\lambda-(n\lambda)^2}$. The
parameter $\nu_0$ here is essentially the same as in
Eq.\eqref{kajhsk}. The normalization factor in front of $K$ is
fixed by matching\ \eqref{lksks} to the asymptotic condition
\eqref{psiassminus}. For $\nu_0\to 0$ and $z\gg 1$ \eqref{lksks}
reduces to\ \eqref{sskshshs}. The Wronskian\ \eqref{wronskian} can
be evaluated in the domain ${1\over n}\ll x-x_0\ll 1$, where both equations\
\eqref{lsj} and \eqref{lksks} are valid; the result is exactly\
\eqref{alasjsl}\ with the extra factor \eqref{shsdy} added.

\subsection{Large $\kappa$ expansion}\label{secseven4}

The case of large $\kappa$ is expected to describe the infrared
limit of the IPH model. For the differential equation\
\eqref{diff}\ it is the domain of validity of the WKB
approximation. Applying the standard WKB iteration scheme\
\cite{Landau}\ one finds for the Wronskian\ \eqref{wronskian},
\bea\label{wkbu} \log W &=& \lim_{L\to+\infty} \bigg\{\, \kappa\,
\int_{-\infty}^{L}\rd x\, \big(\, {\cal P}(x)-\re^{-{nx\over 2}}\,
\big)-\kappa\, \re^{L}-q\, L\, \bigg\} +\nonumber\\  &&
{1\over{8\kappa}}\ \int_{-\infty}^{\infty}\rd x\ {{({\cal
P}'(x))^2} \over{{\cal P}^3(x)}} +\ldots\ , \eea where
\bea\label{kjsks} {\cal P}(x)= \sqrt{\re^{2x}+\re^{-nx}+{2q\over
\kappa}\ \re^x-{p^2\over \kappa^2}}\ . \eea The subtraction term
$\re^{-{nx\over 2}}$ in the integrand derives from the asymptotic
conditions\ \eqref{psiassminus}\ and ensures   convergence of the
first  integral in\ \eqref{wkbu}\ as $x\to-\infty$. The   terms
depending on $L$ cancel  the divergence of this integral as
$x\to+\infty$. It follows from the asymptotic condition
\eqref{psiassplus}. Eq.\eqref{wkbu} generates asymptotic expansion
of the partition function\ \eqref{lksjsa}, \bea\label{expint} \log
Z \simeq\ \log Z_{\rm IR}- \sum_{s=1}^{\infty}\,  {I^{{(\rm
norm)}}_{s}(P,Q)\over 2 \sin({\pi s\over n+2})}\  \Big({n\over
2\kappa}\Big)^s\,, \eea where \bea\label{alkssk} Z_{\rm IR}={\rm
g}_D^2\ r^{-2{\rm i} p}\ \kappa^{-q}\ \  2^{n\, q\over n+2}\ \
{\sqrt{2\pi}\over \Gamma({1\over 2}-q+\ri p)}\ \
\exp\Big(-{\kappa\over \kappa_0}\Big)\ , \eea and the constant
$\kappa_0$ reads explicitly, \bea\label{ksajs}
\kappa_0=-{2\sqrt\pi\over \Gamma(-{n\over 2(n+2)})\Gamma(1-{1\over
n+2})}\ . \eea The functions $I_{s}^{{(\rm norm)}}$ in
\eqref{expint} are polynomials of the variables $P^2$ and $Q$ of
the degree $s+1$. Their highest-order terms follow from the first
integral in\  \eqref{wkbu}, \bea\label{ikhighest}
 I^{{(\rm norm)}}_s&=&{(-1)^{s}\sqrt{\pi}\over 2s}\times\\ &&\sum_{2 l+m=s+1}
{2^{m}(n+2)^{{m\over 2}}\Gamma({1\over 2}-l+{(n+1)s\over n+2})
\over n^{s-l}\, \Gamma(-{s\over n+2})}
  {  P^{2l} Q^m\over m!\, l!}+\ldots\, ,\nonumber
\eea which is in agreement (up to  an overall normalization)
 with the highest-order terms of the vacuum
eigenvalues \eqref{kajhs}. It is also straightforward to generate
the full polynomials evaluating the integral\ \eqref{wkbu}\ order
by order in $\kappa^{-1}$. With the  adjusted overall
normalization of the local IM, this calculation exactly reproduces
the vacuum eigenvalues\ \eqref{kajhs}.

\subsection{$n\to 0$ limit}

There is no any problem with the  limit $n\to 0$  for
the differential equation\ \eqref{diff} and for the
solution $\Psi_+$\ \eqref{psiassplus}.
If  $n=0$ the equation turns to be of  Kummer's type\ \cite{Stegun}  and
\bea\label{manbsmb}
\Psi_+(x)|_{n=0}&=& \re^{-\kappa{\rm e}^x}\
\big(2\kappa\,\re^{x}\big)^{-\sqrt{\kappa^2-p^2}}\times\\ &&
U\big({\textstyle{1\over 2}}+
q-\sqrt{\kappa^2-p^2}, 1-2 \sqrt{\kappa^2-p^2}, 2 \kappa\,
\re^x\, \big)\ .\nonumber
\eea
However, the asymptotic condition\ \eqref{psiassminus} is singular
as $n\to 0$ and
the limiting behavior of the solution $\Psi_-$ is
a more delicate issue. To proceed with the limit
we will use    the WKB approximation for $\Psi_-$:
\bea\label{shksjs}
\Psi_-^{\rm (wkb)}(x)= {1\over \sqrt{2\kappa {\cal P}(x)}}\
\exp\bigg\{\kappa
\, \int_{-\infty}^x\rd t\, \big(\, {\cal P}(t)-\re^{-{nt\over 2}}\, \big)
-{\textstyle{2\kappa\over n}}\
\re^{-{nx\over 2}}\, \bigg\} ,
\eea
where $ {\cal P}(x)$ is given by \eqref{kjsks}.
Let us consider the argument $\{\, \ldots\, \}$ of the
exponential in \eqref{shksjs} as
$n\to 0$. It is easy to see that $\{\, \ldots\, \}\to F(x)-{1\over n}\ S(\kappa,p)$,
where $x$-independent function $S(\kappa,p)$ has the form \eqref{lksjl}, while
$F(x)\to  x\sqrt{\kappa^2-p^2}$ as $x\to+\infty$.
Thus we see that
\bea\label{ssslsl}
\Psi_-(x)\to \Psi^{(\rm reg)}_-(x)\  2^{-{1\over 2}}\ (\kappa^2-p^2)^{-{1\over 4}}\
\ \re^{-{S(\kappa,p)\over n}}\ \ \ \ {\rm as}\ \ \ n\to 0\ ,
\eea
where $\Psi^{(\rm reg)}_-$ is a solution of the differential equation
\eqref{diff} with $n= 0$ such that
$\Psi^{(\rm reg)}_-\to \re^{x\sqrt{\kappa^2-p^2}}$ as $x\to-\infty$.
It is straightforward now to calculate
the boundary amplitude\ \eqref{exactz}. The result coincides
with Eq.\eqref{aoksjo}.

\section{Integrable structures of the theory \label{seceight} }

We  have now seen that the expression\ \eqref{exactz}
passed successfully all available  checks.
In this section
we will   discuss  properties of the vacuum amplitude\ \eqref{exactz}
to reveal  some standard
integrable structures\ \cite{FST,Korep,blz} of the theory.

\subsection{Thermodynamic Bethe Ansatz equations\label{seceight1}}

It is well known from the global theory of linear  ordinary
differential equations\ \cite{Sibuya,Woros} that  monodromic
coefficients of  linear ODE, like the Wronskian\
\eqref{wronskian}, satisfy the difference equations as functions
of parameters. For the equation\ \eqref{diff} with $q=0$ and $n<0$
the corresponding difference equations were derived in the work\
\cite{blz4} (see also \cite{Suzuki,TD}). The case $q=0, \ n>0$ was
studied in the unpublished paper\ \cite{AlZ}.
Here we describe basic properties of the
Wronskian\ \eqref{wronskian} for $n>0$ and
an arbitrary $q$.  For this purpose
it is convenient to modify slightly our   notations. In particular
we introduce  now the parameter $\theta$  such that
\bea\label{slsjksa} \re^{\theta}={\kappa\over \kappa_0}\ , \eea
where the constant $\kappa_0$ is given by\ \eqref{ksajs}. Also, to
emphasize the dependence on $\theta$ and $q={\sqrt{n+2}\over 2}\,
Q$ explicitly, we will denote  the Wronskian\ \eqref{wronskian}\
by ${\rm W}_q(\theta)$. Notice that there is no need to indicate
the dependence on $p={\sqrt{n}\over 2}\, P$ explicitly.

The following  properties of the function ${\rm W}_q(\theta)$ in
the case $n>0$ are readily made using the  general  theory\
\cite{Sibuya} and our previous analysis:

\begin{itemize}

\item  ${\rm W}_q(\theta)$ is   entire  function of  $\theta$.
It is also entire function of the parameters $q$ and $p$.

\item  ${\rm W}_q(\theta)$  satisfies
the so-called ``quantum Wronskian'' condition (see Appendix B for details):
\bea\label{mnmslksjs}
{\rm W}_{-q}\big(\theta-
{\textstyle{{\rm i}\pi\over 2}}\big)
{\rm W}_q\big(\theta+{\textstyle{{\rm i}\pi\over 2}}\big)
-
{\rm W}_q\big(\theta+{\textstyle{{\rm i}\pi a\over 2}}
\big){\rm W}_{-q}\big(\theta-{\textstyle{{\rm i}\pi a\over 2}}\big)
=\re^{-{\rm i}\pi q}\, ,
\eea
where $a={n-2\over n+2}$.

\item  As a function of the complex
variable $\theta$,
${\rm W}_q(\theta)$  does not have zeroes
in the strip
$\big|\Im m\,\theta\big|<{\pi\over 2}+\epsilon$ for some finite $\epsilon>0$.

\item For
$\big|\Im m\,\theta\big|<{\pi\over 2}+\epsilon$ and $\Re e\, \theta\to+\infty$:
\bea\label{ytsalksj}
{\rm W}_q(\theta)= \big(2^{n\over n+2}\, \kappa_0\re^{\theta}\big)^{-q}\
\exp\big(-\re^{\theta}+O(
\re^{-\theta})\, \big)\, .
\eea

\item For $\Re e\, \theta\to-\infty$:
\bea\label{ouyalksj}
{\rm W}_q(\theta)\to L_q(-\ri p)\ \re^{{\rm i} p\theta(n+2)\over n}+
L_q(\ri p)\ \re^{-{{\rm i} p\theta(n+2)\over n}}\ ,
\eea
where
$$L_q(h)=\sqrt{n\over 2\pi}\ \ 2^{-h}\ n^{{2h\over n}}\
(\kappa_0)^{-{(n+2)h\over n}}
\ {\Gamma(2 h)\Gamma(1+{2h\over n})
\over \Gamma({1\over 2}+q+h)}\ .$$

\end{itemize}

\bigskip

Introduce the functions $\varepsilon_{\pm}(\theta)$:
\bea\label{lsks} \re^{\pm {\rm i}\pi q}\ {\rm W}_{q}\big(\theta\pm
{\textstyle{{\rm i}\pi\over 2}}\big)\, {\rm W}_{-q}\big(\theta\mp
{\textstyle{{\rm i}\pi\over
2}}\big)=1+\re^{-\varepsilon_{\pm}(\theta)}\ . \eea With the
foregoing analytical conditions it is straightforward (see
\cite{AlZ1,AlZ2}) to transform the difference equation\
\eqref{mnmslksjs}\ into a system of two  integral equations for
$\varepsilon_{\pm}(\theta)$: \bea\label{ssjahkahk}
&&2\sin\big({\textstyle{2\pi \over n+2}}\big)\,
\re^{\alpha}=\varepsilon_{\pm}(\alpha)\pm {\textstyle {4\pi{\rm i}
q\over n+2}} + \int_{-\infty}^{+\infty}{\rd\beta\over 2\pi}\
\Big\{\, \varphi_{++}(\alpha-\beta)\times\nonumber \\ &&\ \ \ \
\log\big(1 +\re^{-\varepsilon_{\pm}(\beta)}\, \big)+
\varphi_{+-}(\alpha-\beta)\, \log
\big(1+\re^{-\varepsilon_{\mp}(\beta)}\, \big)\, \Big\}\ , \eea
where the kernels are given by $\varphi_{\sigma\sigma'}(\alpha)=-
\ri\ \partial_{\alpha}\log S_{\sigma\sigma'}(\alpha)$ with
\bea\label{slsjs} S_{++}(\alpha)={\sinh({\alpha\over
2}-{{\rm i}\pi\over n+2})\over \sinh({\alpha\over 2}+{{\rm i}\pi\over
n+2})} \, ,\ \ \ \ \ \ \ \ S_{+-}(\alpha)=S_{++}(\ri\pi-\alpha)\ .
\eea The integral equations\ \eqref{ssjahkahk}\ should be
supplemented by an asymptotic condition for
${\varepsilon}_{\pm}(\alpha)$ as $\alpha\to-\infty$ which follows
from\ Eq.\eqref{ouyalksj}. For instance, for $\Im m\, p<0$:
\bea\label{lsjslks}
\varepsilon_{\pm}(\alpha)|_{\alpha\to-\infty}\to\ri p\
{\textstyle{2(n+2)\over n}}\, \ \alpha
 \mp\ri\pi q+\log\big(L_q(\ri p)L_{-q}(\ri p)\big)\ .
\eea
As it follows from Eq.\eqref{lsks}
the Wronskians ${\rm W}_{\pm q}(\theta)$ are expressed in terms of the
solutions of \eqref{ssjahkahk},\,\eqref{lsjslks} by means of   relations:
\bea\label{aslsspoo}
&&\log {\rm W}_{\pm q}(\theta)=
\mp q\log\big(2^{n\over n+2} \kappa_0\re^{\theta}\big)- \re^{\theta}
+{1\over 2}\ \int_{-\infty}^{\infty}{\rd\alpha\over 2\pi}\times\bigg\{
\nonumber \\ &&
{\log\big(1+\re^{-\varepsilon_+(\alpha)}\big)
\log\big(1+\re^{-\varepsilon_-(\alpha)}\big)\over
\cosh(\theta-\alpha)}
\mp\ri
\log\Big({1+\re^{-\varepsilon_+(\alpha)}\over
1+\re^{-\varepsilon_-(\alpha)}}\Big)  {
\re^{\alpha-\theta}\over \cosh(\theta-\alpha)} \bigg\} . \nonumber
\\
\eea It is remarkable that  the system\ \eqref{ssjahkahk} differs
only in the structure of  ``source terms'' from the TBA system
associated with the complex sinh-Gordon model, which is a
non-compact version of the Lund-Regge model\ \cite{Vega,VFat}. It
is well to bear in mind that the classical Lund-Regge model,
introduced in Ref.\cite{Lund}, is a representative of the AKNS
soliton hierarchy. Notice also that if $q=0$, then
$\epsilon_+=\epsilon_-$ and \eqref{ssjahkahk}  turns into  the
integral equation which describes a vacuum boundary amplitude in
the  boundary sinh-Gordon model\ \cite{AlZ}.

\subsection{${\mathbb T}$-operator \label{seceight2} }

Let us consider the differential equation in the form
\eqref{wekajshk} and the solution
$\Psi_-$\ \eqref{psiassminus} as a function of the
variable $y=x+\log(2\kappa)$ and the parameters
$\theta$\ \eqref{slsjksa} and $q$, i.e.,\ $\Psi_-=\Psi_-(\theta,q;y)$.
Notice that  the transformations $\theta\to \theta\pm {2\pi{\rm i}
\over n+2}$ leave ODE\ \eqref{wekajshk} unchanged.
Hence, the functions $\Psi_-(\theta\pm {2\pi {\rm i}
\over n+2},q;y)$ solve this  equation as well as $\Psi_-(\theta,q;y)$, and
the Wronskian
\bea\label{ksspsi}
T(\theta,q)=\ri\ {\rm W}\Big[\,\Psi_-\big(\theta+ {\textstyle{2\pi{\rm i}
\over n+2}},q;y\big)\, ,\, \Psi_-\big(\theta- {\textstyle{2\pi{\rm i}
\over n+2}},q;y\big)\, \Big]
\eea
does not depends on the variable $y$.
It can be shown  (see Appendix B for details)
that function
$T(\theta,q)$ is expressed
through the Wronskian ${\rm W}_q(\theta)$\ \eqref{wronskian}
as
\bea\label{isass}
 {\rm W}_q(\theta)\, T(\theta,q)={\rm W}_q\big(\theta+{\textstyle{2\pi{\rm i}
\over n+2}}\big)+{\rm W}_q\big(\theta-{\textstyle{2\pi{\rm i}
\over n+2}}\big)\ ,
\eea
and
\bea\label{sakjshs}
T\big(\theta+\ri\pi\, {\textstyle{n
\over n+2}}\, ,\, q \big)=T(\theta,-q)\ .
\eea
Since the vacuum amplitude\ \eqref{exactz}
differs by $\theta$-independent factor from
the Wronskian\footnote{Here we  assume that the parameter $r$ in
\eqref{exactz}\ does not depends on $\theta$. This assumption
follows from  the normalization condition
\eqref{majhs}.},
one can replace $W_q$ by $Z$ in \eqref{isass}.
It is also clear
that the function $T(\theta,q)$ admits $\theta\to +\infty$ asymptotic
expansion similar to\ \eqref{expint}.
The foregoing
in turn leads us to

\bigskip

\centerline{\bf CONJECTURE}

\bigskip

\noindent
There exist an  operator ${\mathbb T}$ which  acts invariantly
in the Fock space ${\cal F}_{\bf P}$ and satisfies the following conditions:

\begin{itemize}

\item Being considered as a function of the  parameter
$\lambda\sim\ \re^{\theta (n+2)\over n}$\ \eqref{kajshk},
${\mathbb T}={\mathbb T}(\lambda)$ admits the convergent power
series expansion of the  form: \bea\label{kajhksj} {\mathbb
T}(\lambda)=2\, \cosh\big({\textstyle{\pi P\over \sqrt{n}}}\big)+
\sum_{k=1}^{\infty}{\mathbb G}_k\, \lambda^k\ . \eea The
coefficients ${\mathbb G}_k$ in\ \eqref{kajhksj}  satisfy the
condition, \bea\label{isusasgj} {\mathbb R}\, {\mathbb G}_k\,
{\mathbb R}=(-1)^k\ {\mathbb G}_k\ , \eea where the operator
${\mathbb R}\, :\, {\cal F}_{(P,Q)}\to{\cal F}_{(P,-Q)}$ flips the
overall sign of the field $\partial Y$, i.e., ${\mathbb R}\,
\partial Y\, {\mathbb R}= -\partial Y$.

\item  The operator $ {\mathbb T}(\lambda)$ commutes with all
the IM  from  AKNS series and, hence,
\bea\label{alsjslaaq}
\big[\, {\mathbb T}(\lambda')\, ,\, {\mathbb B}(\lambda)\, \big]=0\ .
\eea
Here the boundary state operator ${\mathbb B}$ is understood as
a  multi-valued function of the spectral parameter\ \eqref{kajshk}.
The eigenvalue of ${\mathbb T}$ corresponding to  the
Fock vacuum $|\, {\bf P}\, \rangle\in{\cal F}_{\bf P}$ coincides with
$T(\theta,q)$\ \eqref{ksspsi}.

\item The boundary state operator ${\mathbb B}$ and ${\mathbb T}$
satisfy the   $T-Q$ Baxter equation\ \eqref{aksks}.

\item The operator ${\mathbb T}(\lambda)$ is a generating
function of the  AKNS series of local IM. In more exact terms
the local IM are generated in the
large $\lambda$ asymptotic series expansion of
${\mathbb T}(\lambda)$:
\bea\label{lsjsjs}
{\mathbb T}(\lambda)=
\exp\Big(-2\pi\, \nu\big(\lambda\, \re^{{{\rm i}\pi\over n}}\big)\, \Big)+
\exp\Big(2\pi\, \nu\big(\lambda\, \re^{-{{\rm i}\pi\over n}}\big)\, \Big)\ ,
\eea
where
\bea\label{alss}
\ri\, {\mathbb \nu}(\lambda)&\simeq &-{\Gamma({n+4\over 2 n+4})\over
\sqrt{\pi}\, \Gamma({n+3\over  n+2})}\, \lambda^{{n\over n+2}}-
{Q\over 2\sqrt{n+2}}+\\ &&{1\over 2\pi}\
\sum_{s=1}^{\infty}\
\Big({R\over 2\lambda^{{n\over n+2}}}\Big)^s\ {\mathbb I}^{{(\rm norm)}}_{s}
\ \ \ \ {\rm as}\ \ \ \  \lambda\to+\infty\ .\nonumber
\eea
The local IM ${\mathbb I}^{{(\rm norm)}}_{s}$ in\ \eqref{alss}  are normalized
in accordance with the condition,
\bea\label{kshksqw}
&&{\mathbb I}^{{(\rm norm)}}_{s}=
{\ri^{s+1}\over n^{s+1\over 2}\, \sqrt{\pi}}\ \  \int_0^{2\pi R}
\rd \tau\, \bigg\{ \ \sum_{2 l+m=s+1}
{\Gamma({1\over 2}-l+{(n+1)s\over n+2})
\over \Gamma(1-{s\over n+2})}\times\nonumber\\
&& 2^{m+s-1}\,
\Big({n+2\over n}\Big)^{{m\over 2}-1}\  {(\partial X)^{2l}\over l!}\,
{(\partial Y)^{m}\over m!}+\ldots\, \bigg\}\ ,
\eea
where omitted terms contain higher derivatives of $\partial{\bf X}$
\footnote{
Note that the local IM ${\mathbb I}^{({\rm norm})}_{s}$ are defined unambiguously
with the normalization condition \eqref{kshksqw}.}.

\end{itemize}

\noindent
Eventually, the operator ${\mathbb T}(\lambda)$ should be
viewed
as a quantum version of the transfer-matrix ${ T}(\lambda)$\ \eqref{kjsjs}
for the AKNS linear problem\ \eqref{alsksl}\ with $j={1\over 2}$.
Following the line of Ref.\cite{blz} it seems possible
to express the coefficients ${\mathbb G}_k$ in the power series expansion
\eqref{kajhksj}
in terms of $2k$-fold integrals over chiral vertex operators
involving the holomorphic component of
field ${\bf X}$. Since the  nonlocal operators ${\mathbb G}_k$
 commute among themselves $[{\mathbb G}_k,{\mathbb G}_m]=0$
and also commute with
all the local IM ${\mathbb I}_s$,  $[{\mathbb G}_k,{\mathbb I}_s]=0$,
they are called the nonlocal Integrals
of Motion. Unfortunately the ``explicit'' formulae  for nonlocal IM
are not particular useful neither to prove the
above listed properties of the operator ${\mathbb T}$ nor
for  calculations of its spectrum.
For this reason we do not present these formulae  here.

Finally we note that  the vacuum eigenvalues $G^{(\rm vac)}_k$ of
nonlocal IM, \bea\label{lkss} {\mathbb G}_k\, |\, {\bf P}\,
\rangle=G^{(\rm vac)}_k\, |\, {\bf P}\, \rangle\ , \eea can be
algebraically expressed in terms of the coefficients
$d_{j}^{(-)}(p,q)$ in the formal power series expansion\
\eqref{skjsu}. For instance,
\bea\label{slksajsajx} G^{(\rm
vac)}_1&=&-4\ n^{n+2\over n}\ \sin\big({\textstyle{\pi\over
n}}\big)\,
\sin\big({\textstyle{\pi(2{\rm i} p +1)\over n}}\big)\ \ d_{1}^{(-)}(p,q)=\\
&&-
{2^{{2\over n}}\,
\Gamma({1\over 2}+{1\over n})\over \sqrt{\pi}
\Gamma(1+{1\over n})}\ \
{8\pi^2\ \ {q\over n}\over \Gamma(
1+{1\over n}-{2{\rm i} p\over n}) \Gamma(
1+{1\over n}+{2{\rm i} p\over n})} \ . \nonumber
\eea

\subsection{Commuting families in the quantum AKNS hierarchy}

In Refs.\cite{blz,blz12} the  quantization procedure for   the KdV
hierarchy was developed. In the BLZ approach  quantum
transfer-matrixes are defined in terms of certain monodromy
matrixes associated with $2j+1$ dimensional representations of the
quantum algebra  $U_{\bf q}\big({\widehat{sl(2)}}\big)$. The
similar $U_{\bf q}\big({\widehat{sl(2)}}\big)$-structure can be
observed in the quantum AKNS hierarchy. In consequence of this the
quantum KdV and AKNS hierarchies share some  common  formal
algebraic properties. In particular,  the quantum
transfer-matrixes ${\mathbb T}_j$ associated with the $2j+1$
dimensional representations of $U_{\bf
q}\big({\widehat{sl(2)}}\big)$ in both hierarchies are recursively
expressed through the operators  ${\mathbb T}\equiv {\mathbb
T}_{1\over 2}$ and the unit operator ${\mathbb I}\equiv{\mathbb
T}_0$ by means of the same fusion relation (see, e.g.,
\cite{blz12}): \bea\label{sskskoiu} {\mathbb T}(\lambda)\,
{\mathbb  T}_j\big({\bf q}^{j+{1\over 2}}\lambda\big)= {\mathbb
T}_{j-{1\over 2}}\big({\bf q}^{j+1}\lambda\big)+{\mathbb
T}_{j+{1\over 2}} \big({\bf q}^{j}\lambda\big)\ . \eea Here  ${\bf
q}$ is the  parameter of  deformation of $U_{\bf
q}\big({\widehat{sl(2)}}\big)$. For the AKNS hierarchy it should
be chosen as \bea\label{slkjsls} {\bf q}=\re^{-{2\pi{\rm i}\over
n}}\ . \eea In Section\,\ref{seceight2} the vacuum eigenvalue of
${\mathbb T}$ was identified with the Wronskian\ \eqref{ksspsi}.
With the fusion relation\ \eqref{sskskoiu} we can express now the
all vacuum eigenvalues, \bea\label{kjshks} {\mathbb
T}_{j}(\lambda)\, |\, {\bf P}\, \rangle=T^{(\rm vac)}_j(\lambda)\,
|\, {\bf P}\, \rangle\ , \eea in terms of solutions of the
differential equation\ \eqref{wekajshk}. The result appears to be
in a remarkable   form generalizing\ \eqref{ksspsi}:
\bea\label{ystssspsi} T^{(\rm vac)}_j(\lambda)= \ri
(-1)^{2j+1}{\rm W}\Big[\Psi_-\big(\theta+ {\textstyle{\pi {\rm
i}(2j+1) \over n+2}},q;y\big) , \Psi_-\big(\theta-
{\textstyle{\pi{\rm i}(2j+1) \over n+2}},q;y\big) \Big]. \eea
Recall that $\lambda=({\kappa_0\over n}\, \re^{\theta})^{n+2\over
n}$,  where the constant $\kappa_0$ is given by \eqref{ksajs}.

We can not resist the temptation to mention here an evidence of
existence of  additional commuting family in the quantum AKNS
hierarchy which does not have a classical counterpart. Indeed, let
us  look at  the solution $\Psi_+$\ \eqref{psiassplus}. On the
complete analogy with the discussion from previous subsection we
consider  $\Psi_+$ as a function of the variable
$y=x+\log(2\kappa)$ and the parameters $\theta$\ \eqref{slsjksa}
and $q$, i.e.,\ $\Psi_+=\Psi_+(\theta,q;y)$. It is easy to see
that the transformation \bea\label{slkssks} y\to y\pm\ri \pi\, ,\
\ \ \theta\to \theta\pm {\textstyle{2\pi{\rm i} n \over n+2}}\, ,\
\ \ \ q\to-q \eea leaves the equation\ \eqref{wekajshk} unchanged.
Because of this the functions
$\Psi_+\big(\theta\pm{\textstyle{{\rm i}\pi n\over n+2}} ,-q;y\pm
\ri\pi \big)$ solve \eqref{wekajshk} and the following  Wronskian
does not depend on  $y$: \bea\label{oeiueuy} {\tilde T}(\theta,q)=
\ri\, {\rm W}\Big[\,\Psi_+\big(\theta-{\textstyle{{\rm i}\pi
n\over n+2}} ,-q;y-\ri\pi \big)\, ,\,
\Psi_+\big(\theta+{\textstyle{{\rm i}\pi n\over n+2}}
,-q;y+\ri\pi\big)\, \Big]\ . \eea It is straightforward to show
that ${\tilde T}(\theta,q)$ satisfies the  relations:
\bea\label{slksjl} W_{q}(\theta)\ {\tilde T}(\theta,q)=
\re^{-{\rm i}\pi q}\ W_{-q}\big(\theta+{\textstyle{{\rm i}\pi n\over
n+2}}\big)+ \re^{{\rm i}\pi q}\ W_{-q}\big(\theta-{\textstyle{{\rm
i}\pi n\over n+2}}\big)\ , \eea and
\bea\label{kashsj} {\tilde
T}\big(\theta-{\textstyle{2{\rm i}\pi\over n+2}}\, ,\, q\big)= {\tilde
T}(\theta,q)\ .
\eea
The evident similarity between\
\eqref{slksjl},\,\eqref{kashsj} and \eqref{ksspsi},\,\eqref{isass}
suggests to interpret ${\tilde T}(\theta,q)$ as a vacuum
eigenvalue of  ``dual'' transfer-matrix ${\tilde {\mathbb T}}$.
Due to Eq.\eqref{kashsj}, it is expected that ${\tilde {\mathbb
T}}$ admits a convergent power series expansion in terms of the
``dual'' spectral parameter ${\tilde \lambda}=\kappa^{n+2}$. Using
properties of the Wronskian\ \eqref{oeiueuy}, it is easy to guess
a set of  general conditions for the operator ${\tilde {\mathbb
T}}({\tilde\lambda})$ which is similar to the one expounded in
Section\,\ref{seceight2} for ${\mathbb T}(\lambda)$. In all
likelihood the appearance of ``dual''  transfer-matrix is related
to the hidden $U_{{\tilde {\bf
q}}}\big({\widehat{sl(2|1)}}\big)$-structure in the quantum AKNS
hirarchy\footnote{The screening operators\
\eqref{comscreen},\,\eqref{kssui} can be interpreted   as
generators of the Borel subalgebra of the  quantum affine
superalgebra $U_{{\tilde {\bf q}}}\big({\widehat{sl(2|1)}}\big)$
with ${\tilde {\bf q}}=\re^{-{{\rm i}\pi n}}$.}.

\section{\label{cocl}Infrared fixed point of the IPH boundary flow}

In this work we  have mainly  focused our attention  on the vacuum
boundary amplitude\ \eqref{lksjsa}. Of course, even the   exact
expression\ \eqref{exactz} does  not define the   boundary state
completely. However,  as we saw in Section\,\ref{seceight}, some
properties of the vacuum eigenvalue inherit important general
features of the whole  boundary state operator. Among them the
remarkable $T-Q$ equation which  is a keystone relation in the
theory of integrable systems. Another property of $Z$ which should
be  understood in more general terms is the large $\kappa$
expansion\ \eqref{expint}. It  suggests that the boundary state
operator ${\mathbb B}$ not only commutes with the local IM, but it
admits an asymptotic expansion in terms of these operators,
\bea\label{ksjhsks} {\mathbb B} \simeq {\mathbb B}_{\rm IR}\
\exp\bigg\{ - \sum_{s=1}^{\infty}\,  {{\mathbb I}^{{(\rm
norm)}}_{s}\over 2 \sin({\pi s\over n+2})}\  \Big({n\over
2E_*}\Big)^s\,\bigg\}\ , \eea where the local IM ${\mathbb
I}^{{(\rm norm)}}_{s}$ are normalized in accordance with the
condition\ \eqref{kshksqw}.

From the physical point of view the
most interesting object  to be discussed is the infrared boundary
state associated with  the operator ${\mathbb B}_{\rm IR}$\ \eqref{ksjhsks}. According to  our
consideration this  boundary   state  should have the form
\bea\label{kshssh}
|\, B\, \rangle_{\rm IR}=\int_{ {\bf P}}
\rd^2{\bf P}\ Z^*_{\rm IR}\ |\, {\cal I}_{\bf P}\, \rangle\ ,
\eea
where the    amplitude $Z_{\rm IR}$ is given by
Eq.\eqref{alkssk}  and therefore
the
states $|\, {\cal I}_{\bf P}\, \rangle\in {\cal F}_{\bf P}\otimes {\bar {\cal F}}_{\bf P}$
are normalized as follows:
\bea\label{klsusa}
\langle\, {\bf P}'\, |\,  {\cal I}_{\bf P}\, \rangle=\delta({\bf P}'-{\bf P})\ .
\eea
Of course,
the   boundary state \eqref{kshssh}\ should obey the integrability condition  \eqref{constrW}.
It is  also  expected
to be a {\it conformal} boundary state
which is some   deformation to the domain $n>0$ of
the  infrared  boundary state with $n=0$ described
in Section\, \ref{secsix2}.
In Appendix C
it is shown that
a boundary state satisfying the above mentioned conditions
must also  possess an extended conformal  symmetry
generated by
the holomorphic  spin-1 and spin-2 currents:
\bea\label{alasjs}
\big[\, W^{(\rm IR)}_{s}(\tau)-
{\bar W}^{(\rm IR)}_{s}(\tau)\, \big]_{\sigma=0}\, |\, B\, \rangle_{\rm
IR}=0\ \ \ \ \ \ \ (s=1,\, 2)\  ,\eea
where
\bea\label{akshksah}
W^{(\rm IR)}_1&=&\partial { X}^\star\ ,\nonumber\\
W^{(\rm IR)}_2&=&-(\partial { Y}^\star)^2+{\rm i\over \sqrt{2}}\ \partial^2 { Y}^\star\ .
\eea
In Eq.\eqref{akshksah}\ we use the notations
\bea\label{salkhj}
{ X}^\star={\sqrt{n+2\over 2}}\ X-{\rm i}\ {\sqrt{n\over 2}}\ Y\, ,
\ \ \ { Y}^\star={\sqrt{n+2\over 2}}\ Y+{\rm i}\ {\sqrt{n\over 2}}\ X\ .
\eea
Notice   that the  spin-2 current $W^{\rm (IR)}_2$ produces the Virasoro
algebra with  center charge $c=-2$.
If
\bea\label{assksl}
{ q}^\star={\sqrt{n+2}\over 2}\ Q+{\rm i}\ {\sqrt{n} \over 2}\ P\not=\ \pm
{{1\over 2}}\,,\,
\pm {{3\over 2}}\,,\ldots\ ,
\eea
the building blocks (Ishibashi states)   $|\, {\cal I}_{\bf P}\, \rangle$ in\ \eqref{kshssh}
are defined by Eqs.\eqref{klsusa}, \eqref{alasjs} uniquely for
the Fock spaces
${\cal F}_{\bf P}\otimes{\bar {\cal F}}_{\bf P}$ with the zero-mode momentum
${\bf P}=(P,Q)$\ \cite{Ishibashi, Cardy}.
In particular,
\bea\label{jhsksh}
|\, {\cal I}_{\bf P}\, \rangle&=&\exp
\Big(\sum_{k=1}^{\infty} {\textstyle {2\over k}}\, {X}^\star_{-k}
{ {\bar X}^\star}_{-k}\, \Big)\
 \bigg[\, 1+ { { q}^\star+{1\over 2}\over { q}^\star
-{1\over 2}}\ \ 2\, { Y}^\star_{-1}{ {\bar Y}}^\star_{-1}+\nonumber\\ &&
{ { q}^\star+{3\over 2}\over { q}^\star
-{1\over 2}}\ \big(\, { (Y^\star_{-1})}^2+{\textstyle{1\over \sqrt{2}}}\,
{ Y}^\star_{-2}
\ \big)\big(\,   ({\bar Y}^\star_{-1})^2+
{\textstyle{1\over \sqrt{2}}}\,{\bar Y}^\star_{-2}\
\big)+ \\ &&
{ { q}^\star+{1\over 2}\over { q}^\star
-{3\over 2}}\
\big(\,  (Y^\star_{-1})^2-{\textstyle{1\over \sqrt{2}}}\,
{ Y}^\star_{-2}
\ \big)\big(\,   ({\bar Y}^\star_{-1})^2-
{\textstyle{1\over \sqrt{2}}}\,{\bar Y}^\star_{-2}\
\big)+\ldots\,
\bigg]\  |\, {\bf P}\, \rangle\ .
 \nonumber
\eea
Here ${ X}^\star_{-k},\, { Y}^\star_{-k}$ are the oscillatory modes of
the fields  \eqref{salkhj}.

It can be shown that the integrand in Eq.\eqref{kshssh}, where $|\, {\cal I}_{\bf P}\, \rangle$
given by \eqref{jhsksh}, is
well defined even for the exceptional values
${ q}^\star=\pm{1\over 2},\, \pm{3\over 2}\ldots$ and
therefore,
we propose  \eqref{kshssh},\,\eqref{jhsksh}\
for  the infrared  boundary state of   the IPH model.
Notice that it splits into  the Dirichlet boundary state corresponding to the Dirichlet
boundary condition $X^\star_{B}=const$,
and the $n$-independent conformal   boundary state  for the field  ${ Y}^\star$.
Unfortunately a consistent   Lagrangian description
of the   ${ Y}^\star$-component of $|\, B\, \rangle_{\rm IR}$
is still a question for the authors.

We would like to  conclude the paper by a remark about
the higher level boundary amplitudes $B_{\alpha}({\bf P})$ in Eq.\eqref{ssshsaksa}.
It seems likely
that they also
admit a
description in terms of ODEs similar to\ \eqref{diff}. Such
higher level differential equations can be constructed
along the line of Ref.\cite{blz8} exploring
analytical properties of the boundary state operator ${\mathbb B}$ as a function of
variable $\theta$\ \eqref{slsjksa}.
We intend to return to this problem in a future work.

\section*{Acknowledgments}

This work was started during SLL visit at SdPT at Saclay and LPTA
Universit${\acute{\rm e}}$ Montpellier II in April 2005. He is
grateful to members of these laboratories and especially to Ivan
K. Kostov for their kind hospitality  and interesting discussions.
The authors are  also grateful to Alexander and Alexei B. Zamolodchikov
for sharing their insights and important comments.
\bigskip

\noindent The research of VAF is supported by the grant
INTAS-OPEN-03-51-3350 and by the EU under contract EUCLID
HRPN-CT-2002-00325.

\noindent The research of SLL is supported in part by DOE grant
$\#$DE-FG02-96 ER 40949.

\bigskip
\bigskip


\section{Appendix A}\label{usalksj}

Here we present an explicit form of the first local IM of the AKNS
series as an operators acting in the Fock space. Let us introduce
the notations, \bea\label{lksajs} {\mathbb X}_{i_1\ldots
i_m}^{\mu_1\ldots \mu_m}=\sum_{s_1+\ldots s_m=0\atop
s_j\not=0}(s_1)^{i_1}\ldots (s_m)^{i_m}\ :X^{\mu_1}_{s_1}\ldots
X^{\mu_m}_{s_m}:\ . \eea The normal ordering $:\ :$ in this
formula means that the operators $X^{i}_s$\ \eqref{sakshs} with
the bigger $s$ are placed to the right. The local IM can be
written as follows: \bea\label{kjhdskjd} {\mathbb I}_1&=&R^{-1}\
\big[\,
{\mathbb X}^{11}_{00}+{\mathbb X}^{22}_{00}+I^{({\rm vac})}_1(P,Q)\, \big]\,,\\
{\mathbb I}_2&=&
R^{-2}\
\big[\, {\textstyle{6n+4\over 3}}\,
{\mathbb X}^{222}_{000}+2n\, {\mathbb X}^{112}_{000}+
(3n+2)\, Q\, {\mathbb X}^{22}_{00}+
n\, Q\, {\mathbb X}^{11}_{00}+\nonumber
\\ &&
2n\, P\, {\mathbb X}^{12}_{00}-2\ri\,(n+1)\, \sqrt{n}\,
 {\mathbb X}^{12}_{10}+I_2^{({\rm vac})}(P,Q)\, \big]\, ,
\nonumber
\eea
and
\bea\label{basvs}
{\mathbb I}_3 &=&R^{-3}\ \big[\,
n\, {\mathbb X}^{1111}_{0000}+(5n+4)\,
{\mathbb X}^{2222}_{0000}+6n\, {\mathbb X}^{1122}_{0000}+
2Pn\, {\mathbb X}^{111}_{000}+\nonumber \\ &&
2 (5n+4)\, Q\, {\mathbb X}^{222}_{000}
+6nP\, {\mathbb X}^{122}_{000}+6nQ\, {\mathbb X}^{112}_{000}-
6\ri\, \sqrt{n}\, (n+1) \, {\mathbb X}^{122}_{100}+\nonumber \\ &&
{\textstyle{ n\over 2}}\, ( 3P^2+3Q^2-1)\, {\mathbb X}^{11}_{00}+
{\textstyle{ 1\over 2}}\,
 (3(5n+4)Q^2+ 3nP^2-3n-2)\, {\mathbb X}^{22}_{00}+\nonumber \\ &&
6n PQ\, {\mathbb X}^{12}_{00}-
6\ri\, \sqrt{n}\, (n+1)\, Q\ {\mathbb X}^{12}_{10}-\nonumber\\ &&
(n^2+3 n+1)\, {\mathbb X}^{11}_{11}
-(n^2+4 n+2)\, {\mathbb X}^{22}_{11}+I^{({\rm vac})}_3(P,Q)\, \big]\, .
\eea
Here $I^{({\rm vac})}_s(P,Q)$ are
the vacuum eigenvalues of the AKNS integrals\ \eqref{skkjashk}.

\section{Appendix B}\label{ieualksj}

Here we  prove the quantum Wronskian condition\ \eqref{mnmslksjs} and
Eq.\ \eqref{isass}.

We start with an observation that the following transformations
of the variables $(y,\kappa,q)$,
\bea\label{muynmsjs}
{\hat \Lambda}&:&\ y\to y+\ri\pi\, ,\ \ \ \kappa\to\re^{{\rm i}\pi n\over n+2}
\, \kappa,\ \ \  q\to -q\ ,\nonumber\\
{\hat \Omega}&:&\ y\to y\, ,\ \ \ \kappa\to\re^{2{\rm i}\pi\over
n+2},\ \ \  q\to q\ , \eea leave the ODE \eqref{wekajshk}
unchanged while acting nontrivially on its solutions. The
transformation ${\hat \Lambda}$ applied to the solution $\Psi_+$\
\eqref{psiassplus} yields another solution, and the pair of
functions \bea\label{lksjssauj} \Psi_+=\Psi_+(y,\theta,q)\, ,\ \ \
{\hat \Lambda}\Psi_+=\Psi_+\big(y+\ri\pi,\theta+{\textstyle{{{\rm
i}\pi n\over n+2}}},-q\big)\ , \eea with $\theta$ given by\
\eqref{slsjksa}, forms a basis in the space of solutions of\
\eqref{wekajshk}. It is not difficult to check that,
\bea\label{mnasajsla} W\big[\Psi_+,{\hat
\Lambda}\Psi_+\big]=\re^{{\rm i}\pi(q-{1\over 2})}\ , \eea i.e.,
the solutions\ \eqref{lksjssauj}\ are indeed linearly independent.
The solution $\Psi_-$\ \eqref{psiassminus} can always be expanded
in this basis, in particular \bea\label{askss} \Psi_-=a\ \Psi_++b\
{\hat \Lambda}\Psi_+\ . \eea Using Eq.\eqref{mnasajsla} we
conclude that \bea\label{lakjsla} b=\re^{{\rm i\pi}({1\over
2}-q)}\ W_q(\theta)\ , \eea where
$W_q(\theta)=W\big[\Psi_+,\Psi_-\big]$. The transformation ${\hat
\Lambda}$ leaves the solution $\Psi_-$ unchanged\footnote{This property
of  $\Psi_-$ leads immediately to Eq.\eqref{sakjshs}.}, i.e.,
$\Psi_-(y,\theta,q)=\Psi_-\big(y+\ri\pi,\theta+{\textstyle{{{\rm
i}\pi n\over n+2}}},-q\big)$. This allows one to express the
coefficient $a$ in \eqref{askss}    in terms of the Wronskian
$W_q(\theta)$ as well: \bea\label{uddylakjsla} a=- \re^{{\rm
i\pi}({1\over 2}-q)}\ W_{-q}\big(\theta+{\textstyle {{\rm i}\pi
n\over n+2}}\big)\ . \eea Let us apply now the transformation
${\hat \Omega}$\ \eqref{muynmsjs} to the both sides of
Eq.\eqref{askss}. It is apparent that $\Psi_+$ is invariant with
respect to the action of ${\hat \Omega}$, and hence,
\bea\label{lsajsa} {\hat \Omega}\Psi_-=\re^{{\rm i\pi}({1\over
2}-q)}\ \Big(\,W_q\big(\theta+{\textstyle {2{\rm i}\pi \over
n+2}}\big) \ {\hat \Lambda}\Psi_+ - W_{-q}(\theta+\ri\pi)\
\Psi_+\, \Big)\ . \eea The quantum Wronskian condition
\eqref{mnmslksjs} follows immediately
 from \eqref{askss},\ \eqref{lsajsa},\ \eqref{mnasajsla} and
the relation
\bea\label{muyasajsla}
W\big[\Psi_-,{\hat \Omega}\Psi_-\big]=\ri\ .
\eea

It is easy to derive now from the last equation that three
solutions $\Psi_-$,\ ${\hat \Omega}\Psi_-$ and ${\hat
\Omega}^{-1}\Psi_-$ satisfy the relation,
\bea\label{muyasajslab}
T(\theta,q)\,\Psi_-={\hat \Omega}\Psi_- +{\hat \Omega}^{-1}\Psi_-\ ,
\eea 
with function $ T(\theta,q)$ defined by  Eq.\,\eqref{ksspsi}.
Taking the Wronskian from the both sides of this equation with the
solution $\Psi_+ ={\hat \Omega}\Psi_+ ={\hat \Omega}^{-1}\Psi_+$
we arrive to Eq.\,\eqref{isass}.

\section{Appendix C}\label{ilksj}

Here we show that the infrared boundary state of the IPH model,
$|\, B\, \rangle_{\rm IR}$,
satisfies the conditions\ \eqref{alasjs}-\eqref{salkhj}.

Our analyses is based on the assumption that $|\, B\, \rangle_{\rm IR}$  possesses the conformal
symmetry, i.e.:
\bea\label{klsahak}
\big[\, T^{(\rm IR)}(\tau)-{\bar T}^{(\rm IR)}(\tau)\, \big]_{\sigma=0}\, |\, B\, \rangle_{\rm IR}=0\ ,
\eea
where the holomorphic and antiholomorphic components of
infrared stress-energy tensor
have the  most general admissible  form:
\bea\label{jhgs}
T^{(\rm IR)}&=&-\partial {\bf X}\cdot \partial{\bf X}+\ri\, {\boldsymbol \rho}\cdot\partial^2 {\bf X}\ ,\\
{\bar T}^{(\rm IR)}&=&-{\bar \partial} {\bf X}\cdot {\bar \partial}
{\bf X}+\ri\, {\boldsymbol \rho}\cdot{\bar \partial}^2 {\bf X}\ .\nonumber
\eea
The  constant vector ${\boldsymbol \rho}=(\rho_x,\rho_y)$ is unknown a priori, but  it is expected to be
a function of coupling constant $n$.
A comparison between\ \eqref{jhgs}\ and
the infrared stress-energy tensor for $n=0$\ \eqref{alksjo} shows that
\bea\label{lasjl}
\rho_y|_{n=0}={\textstyle {1\over \sqrt{2}}}\ .
\eea

The  local fields \eqref{jhgs}  and the first nontrivial
local IM ${\mathbb I}_2\ ({\bar {\mathbb I}}_2)$\ \eqref{kjhdskjd} allow one to
construct   holomorphic
spin-3 and antiholomorphic spin-(-3)  currents through the relations:
\bea\label{jsahks}
\partial W_3^{\rm(IR)}= [\, T\, ,\, {\mathbb I}_2\, \big]\ ,\ \ \ \
{\bar \partial}
{\bar W}_3^{\rm(IR)}= [\, {\bar T}\, ,\, {\bar {\mathbb I}}_2\, \big]\ .
\eea
Since the infrared boundary state
satisfies both\  Eq.\eqref{klsahak}\ and
the integrability condition \eqref{alksla}, it is apparent that
\bea\label{sahak}
\big[\, W^{(\rm IR)}_3(\tau)-
{\bar W}_3^{(\rm IR)}(\tau)\, \big]_{\sigma=0}\, |\, B\, \rangle_{\rm IR}=0\ .
\eea
An explicit form of $W^{(\rm IR)}_3$ is not particularly  important for the
present discussion. It is important  that this field and $T^{(\rm IR)}$
generate a spin-1 current $W^{(\rm IR)}_1$ through their operator product expansion:
\bea\label{skssl}
T^{\rm (IR)}(u) W^{\rm (IR)}_3(v)=
{6W^{\rm (IR)}_1(v)\over 
(u-v)^4}+{2\partial W^{\rm (IR)}_1(v)\over (u-v)^3}+O\big((u-v)^{-2}\big)\ .
\eea
Explicit calculations
shows that  $W^{\rm (IR)}_1$ has the form
\bea\label{sytrlksjl}
W^{\rm (IR)}_1=\alpha\, \partial X+\ri\, \beta\, \partial Y\ ,
\eea
with
\bea\label{sakjsahas}
\alpha&=&2\, \rho_y \sqrt{n}\ (1+n +  {\rm i} \rho_x \sqrt{n} )\, ,\\
\beta&=& n\, \rho_x^2+\rho_y^2\,  (2 + 3 n)
+ 2 {\rm i}\, \rho_x\, (1+n) \sqrt{n}-2n-1 \ .\nonumber
\eea
As it follows from Eqs.\eqref{sakjsahas}\ the current $W^{\rm (IR)}_1$ may  vanish
for $\rho_y=0$ or $\rho_y=\pm\sqrt{n+2}$ only. Since none  of these  is  consistent
with Eq.\eqref{lasjl}, the infrared boundary state  should satisfy
the condition\ \eqref{alasjs} for the nonvanishing spin-1 current
given by Eqs.\eqref{sytrlksjl} and \eqref{sakjsahas}.

Let us consider now the operator product expansion $W^{\rm (IR)}_1(u)W_3^{\rm (IR)}(v)$.
It contains a singular term $\sim (u-v)^{-2}$ which
involves the  spin-2 current of the form,
\bea\label{lkslsjkj}
{\tilde W}^{\rm (IR)}_2=
2\ri n\beta (\partial X)^2+4n\alpha \partial X\partial Y+\ri
\beta (6n+4) (\partial Y)^2+
\gamma\partial^2X+\delta\partial^2Y ,
\eea
where
\bea\label{uklsjsl}
\gamma&=& n\, (\beta\rho_x-\ri\alpha\rho_y)+
\ri\, (n+1)\, \beta\, \sqrt{n}\ ,\\
\delta&=&-\alpha\, \sqrt{n}\, (n+1+\ri\rho_x\sqrt{n}\, )+\beta\, \rho_y\, (3n+2)\ .\nonumber
\eea
With Eqs.\eqref{sytrlksjl},\,\eqref{lkslsjkj} it is easy to see
that   the operator product expansion
$W^{\rm (IR)}_1(u){\tilde W}_2^{\rm (IR)}(v)$ produces
one more
spin-1 current
\bea\label{trwwskjskj}
{{\tilde W}^{\rm (IR)}_1}=2 n{\rm i}\, \alpha\beta\ \partial X+
(\, n\alpha^2-(3n+2)\, \beta^2\,)\ \partial Y\ .
\eea
This  current does not vanish provided  ${ W}^{\rm (IR)}_1$ exists.
Therefore one needs to explore
two possibilities: either ${ W}^{\rm (IR)}_1$ and ${\tilde W}^{\rm (IR)}_1$
are
linear independent or  linear dependent currents.
The first possibility  implies that $|\, B\, \rangle_{\rm IR}$
is  the  $n$-independent  Dirichlet boundary state
associated with the Dirichlet   boundary condition  ${\bf X}_B={\bf const}$.
It should be  apparently ignored.
Hence   ${\tilde W}^{\rm (IR)}_1\sim W^{\rm (IR)}_1$.
With Eqs.\eqref{sytrlksjl},\,\eqref{trwwskjskj} it gives the relation
\bea\label{iusksahsa}
\sqrt{n}\ \alpha\pm \sqrt{n+2}\ \beta=0\ .
\eea
Now it is convenient  to introduce
a new set $( X^\star,\,  Y^\star)$
related to the basic fields $(X,\, Y)$ through the (complex) orthogonal transformation:
\bea\label{osuisalkhj}
 X^\star={\sqrt{n+2\over 2}}\ X\mp {\rm i}\ { \sqrt{n\over 2}}\ Y\, ,
\ \ \  Y^\star= \ri\ { {\sqrt{n\over 2}}}\ X\pm \
{{\sqrt{n+2\over 2}}}\ Y\ .
\eea
Here the sign factors
are  dictated by the choice of sign in \eqref{iusksahsa}.
Since $W_1^{\rm (IR)}\sim \partial X^\star$ one can choose (without loss of generality)
the infrared
stress-energy  tensor\ \eqref{jhgs}\ in the form
\bea\label{sshlp}
T^{\rm (IR)}=-(\partial X^\star)^2+W_2^{\rm (IR)}\ ,
\eea
where
\bea\label{aisah}
W_2^{\rm (IR)}=-(\partial  Y^\star)^2+{{\rm i}\sigma
\over \sqrt{2}} \ \partial^2  Y^\star\ .
\eea
It follows immediately from Eqs.\eqref{sakjsahas} and  \eqref{iusksahsa} that  $\sigma^2=1 $.

Thus  there are
four nonvanishing
spin-2 currents,
$T^{\rm (IR)}$,\ $W_2^{\rm (IR)}$, ${\tilde W}^{\rm (IR)}_2$
and $\partial^2{ X}^\star$,
satisfying the condition $(W_2-{\bar W}_2)\, |\, B\,\rangle_{\rm IR}=0$.
If one admits now  their   linear independence,
then  $\partial^2  Y^\star$ can be expressed in terms of these currents
and, consequently,  should also satisfy the above condition.
This  will eventually lead  us
to the  Dirichlet boundary state  corresponding to
${\bf X}_B={\bf const}$. Therefore we reject this possibility and
accept that the current ${\tilde W}^{\rm (IR)}_2$ is
linearly expressed in
terms of $T^{\rm (IR)}$,\ $W_2^{\rm (IR)}$ and $\partial^2 X^\star$.
It is straightforward to  check that this is
indeed the case
provided
\bea\label{slksjs}
\sigma=+1\ .
\eea
The remaining ambiguity in   sign factors
in Eqs.\eqref{osuisalkhj}\ should be resolved  in accordance
with the condition\ \eqref{lasjl}.
In  this way one arrives to Eqs.\eqref{alasjs}-\eqref{salkhj}.

Finally we note that the
local IM ${\mathbb I}_2$\ \eqref{kjhdskjd}\  can be expressed
in terms of the fields\ \eqref{akshksah},\,\eqref{salkhj}  as follows
\bea\label{jkagsj}
{\mathbb I}_2=(2n+1)\sqrt{n+2\over 2}\ {\mathbb I}^{(n=0)}_2[\,{ Y}^\star\,]
-2\sqrt{2n}\ {\mathbb J}_2\big[\, W_1^{\rm (IR)},\, W_2^{\rm (IR)}\,\big]\ ,
\eea
where
\bea\label{asajso}
{\mathbb I}^{(n=0)}_2[\,  Y^\star\, ]={4\ri \over 3}\
\int_{0}^{2\pi R}{\rd \tau\over 2\pi}\ (\partial  Y^\star)^3\ ,
\eea
and
\bea\label{kshks}
{\mathbb J}_2[W_1,W_2]=\int_{0}^{2\pi R}{\rd \tau\over 2\pi}\ \big(\, (n+1)\
W_2W_1+{\textstyle{2n+3\over 3}}\ (W_1)^3\, \big)\ .
\eea
Evidently $({\mathbb J}_2-{\bar {\mathbb J}}_2)\, |\, B\, \rangle_{\rm IR}=0$,
therefore  the integrability condition
$({\mathbb I}_2-{\bar {\mathbb I}}_{-2})\, |\, B\, \rangle_{\rm IR}=0$
for the  boundary state\ \eqref{kshssh},\,\eqref{jhsksh} with $n> 0$
follows from the fact that it holds for  $n=0$.

\end{document}